\PassOptionsToPackage{dvipsnames}{xcolor}

\documentclass[pmlr]{jmlr}

\RequirePackage{graphicx}
 \usepackage{booktabs}
\usepackage{longtable}%

\makeatletter
\def\set@curr@file#1{\def\@curr@file{#1}} %
\makeatother
\usepackage[load-configurations=version-1]{siunitx} %

\theorembodyfont{\upshape}
\theoremheaderfont{\scshape}
\theorempostheader{:}
\theoremsep{\newline}

\jmlrvolume{298}
\jmlryear{2025}
\jmlrworkshop{Machine Learning for Healthcare}

\usepackage[dvipsnames]{xcolor}
\definecolor{myblue}{RGB}{31, 119, 180}
\definecolor{myorange}{RGB}{255, 127, 14}
\definecolor{mygreen}{RGB}{44, 160, 44}
\definecolor{myred}{RGB}{214, 39, 40}
\definecolor{myyellow}{RGB}{230,194,0}

\usepackage{algorithm}
\usepackage{algpseudocode}

\usepackage{cite}
\usepackage{comment}
\usepackage{prettyref}
\usepackage{amsfonts}
\usepackage{makecell}
\usepackage{adjustbox}
\usepackage{multicol}
\usepackage{wrapfig}

\usepackage{subcaption}

\usepackage{amsmath}
\usepackage{amssymb}
\usepackage{mathtools}

\newcommand{\equal}[1]{{\hypersetup{linkcolor=black}\thanks{#1}}}

\usepackage{amssymb}
\usepackage{mathtools}

\usepackage{xspace}
\usepackage{enumitem}

\definecolor{colorcomment}{RGB}{160, 190, 210}%
\makeatletter
\algnewcommand{\LineComment}[1]{\Statex \hskip\ALG@thistlm \(\triangleright\) 
{\color{colorcomment}#1}}
\makeatother

\makeatletter
\algnewcommand{\IndentLineComment}[1]{\Statex \hskip\ALG@tlm \(\triangleright\) {\color{colorcomment}#1}}
\makeatother

\newcommand\policy{\ensuremath{\pi}}

\newcommand\state{s}

\newcommand\stateDist{d}

\newcommand\transDynamics{\mathcal{P}}

\newcommand\RFunc{R}

\newcommand\Pair{\mathcal{P}}

\newcommand\gradient{{g}}

\newcommand\VFunc{\ensuremath{\ensuremath{V}}}

\newcommand\QFunc{\ensuremath{\ensuremath{Q}}}

\newcommand{\given}{{\,|\,}}
\newcommand{\Prob}{{P}}

\newcommand{\AP}{\mathtt{AP}}

\newcommand\stateSpace{\ensuremath{\mathcal{S}}}
\newcommand\action{\ensuremath{a}}

\newcommand\actionSpace{\ensuremath{\mathcal{A}}}

\newcommand\horizon{\ensuremath{T}}
\newcommand\ENum{\ensuremath{N}} %
\newcommand\corpus{\ensuremath{C}}

\newcommand\critic{\ensuremath{{C}}\xspace}

\newcommand\MDP{\ensuremath{\mathcal{M}}}

\newcommand{\expctover}[2]{\mathbb{E}_{#1}\!\left[#2\right]}
\newcommand\RationaleBuffer{\ensuremath{\mathcal{B}}}

\newcommand\BON{\text{{{BON}}}}

\def \argmax {\mathop{\rm arg\,max}}
\def \argmin {\mathop{\rm arg\,min}}

\newcommand\Gen{\ensuremath{\policy_{{\theta}}}}

\newcommand{\TOPK}{\textsc{Top-K}\xspace}
\newcommand{\TOPP}{\textsc{Top-P}\xspace}
\newcommand{\TOPN}{\textsc{Top-N}\xspace}
\newcommand{\TOPPK}{\textsc{Top-PK}\xspace}
\newcommand{\hypothesis}{\mathcal{Y}}

\newcommand\Buffer{\ensuremath{\mathcal{B}}}

\newcommand{\algname}{\textsc{ScaffoldGPT}\xspace}

\newif\iffinal
\finaltrue

\iffinal
    \newcommand{\fix}[1]{#1}
    \newcommand{\pref}[1]{}
    \newcommand{\XL}[1]{}
    \newcommand{\SJ}[1]{}
    \newcommand{\XLinline}[1]{}
\else
    \newcommand{\fix}[1]{{\color{red} #1}}
    \newcommand{\SJ}[1]{\todo[fancyline,color=Blue!40]{SJ: #1}\xspace}
    
    \newcommand{\XL}[1]{\todo[fancyline,color=Maroon!40]{XL: #1}\xspace}
    \newcommand{\XLinline}[1]{\textcolor{Maroon}{[XL: #1]}}

    \newcommand{\pref}[1]{{\color{blue}(\ref{#1})}}
\fi

\newcommand{\tabref}[1]{Table~\ref{#1}}
\newcommand{\figref}[1]{Fig.~\ref{#1}}

\newcommand{\secref}[1]{\S\ref{#1}}
\newcommand{\appref}[1]{Appendix~\ref{#1}}

\newcommand{\paren} [1] {\ensuremath{ \left( {#1} \right) }}

\newcommand{\bracket}[1]{\left[#1\right]}
\newcommand{\tuple}[1]{\ensuremath{\left\langle #1 \right\rangle}}
\newcommand{\curlybracket}[1]{\ensuremath{\left\{#1\right\}}}

\title[\algname]{\algname: A Scaffold-based GPT Model for Drug Optimization}

\author{\Name{Xuefeng Liu}\equal{These authors contributed equally.\\ X. Liu, I. Foster, and R. Stevens are also affiliated with Argonne National Laboratory.} 
       \Email{xuefeng@uchicago.edu}\\ 
       \addr Department of Computer Science\\
       University of Chicago\\
       Chicago, IL, U.S. 
       \AND
       \Name{Songhao Jiang}\footnotemark[1]
       \Email{shjiang@uchicago.edu}\\ 
       \addr Department of Computer Science\\
       University of Chicago\\
       Chicago, IL, U.S. 
        \AND
       \Name{Ian Foster}
       \Email{foster@uchicago.edu }\\ 
       \addr Department of Computer Science\\
       University of Chicago\\
       Chicago, IL, U.S. 
        \AND
        \Name{Jinbo Xu}
       \Email{jinboxu@gmail.com}\\ 
       \addr Toyota Technological Institute at Chicago\\
       Chicago, IL, U.S.  
        \AND
        \Name{Rick Stevens}
       \Email{stevens@cs.uchicago.edu}\\ 
       \addr Department of Computer Science\\
       University of Chicago\\
       Chicago, IL, U.S. 
       }

\begin{document}

\maketitle

\begin{abstract}

Drug optimization has become increasingly crucial in light of fast-mutating virus strains and drug-resistant cancer cells. Nevertheless, it remains challenging as it necessitates retaining the beneficial properties of the original drug while simultaneously enhancing desired attributes beyond its scope. 
In this work, we aim to tackle this challenge by introducing \algname, a novel \fix{Generative Pretrained Transformer (GPT)} designed for drug optimization based on molecular scaffolds. Our work comprises three key components:
(1) A three-stage drug optimization approach that integrates pretraining, finetuning, and decoding optimization.
(2) A novel two-phase incremental pre-training strategy for scaffold-based drug optimization.
(3) A token-level decoding optimization strategy, \TOPN, that enabling controlled, reward-guided generation using the pretrained or finetuned GPT.
We demonstrate via a comprehensive evaluation on COVID and cancer benchmarks that \algname outperforms the competing baselines in drug optimization benchmarks, while excelling in preserving original functional scaffold and enhancing  desired properties.

\end{abstract}

\section{Introduction}\label{sec:intro}

The rise of rapidly mutating virus strains~\citep{hadj2022covid}, exemplified by those related to SARS-CoV-2~\citep{yuki2020covid}, along with drug-resistant cancer cells~\citep{MANS2017}, has heightened the urgency and interest in accelerating the development of effective treatments. 
However, traditional De Novo drug discovery processes are prohibitively expensive, often costing from hundreds of millions to billions of dollars~\citep{dickson2009cost}, due to their extensive and resource-demanding nature. 
Despite considerable efforts, the success rate of drug discovery remains low, with many candidates failing early in the development process.
This has led to an increased focus on drug repurposing, which leverages existing FDA-approved drugs for new therapeutic uses rather than creating new drugs from scratch. While drug repurposing has seen some success~\citep{pushpakom2019drug}, its effectiveness is often constrained because drugs are typically developed with a high specificity for a particular disease.

Drug optimization seeks to overcome the limitations inherent in both De Novo drug discovery and drug repurposing by enhancing an existing FDA-approved drug.  
Drug optimization is becoming an increasingly vital field, yet it remains relatively underexplored compared to drug discovery and repurposing efforts.
DrugImprover~\citep{liu2023drugimprover} has started to clearly define the drug optimization problem by using Tanimoto similarity. 
Additionally, the DrugImprover framework contributes to the drug optimization domain in three key aspects: a detailed workflow for drug optimization, an Advantage-alignment Policy Optimization (APO) reinforcement learning (RL) algorithm to enhance the multi-objective generative model for drug optimization, and an extensive dataset featuring 1 million ligands and their OEDOCK scores for five proteins related to human cancer cells and 24 high-affinity binding sites on the SARS-CoV-2 protein 3CLPro (PDB ID: 7BQY). The DrugImprover framework features a pretrained generative model that is subsequently refined using the APO reinforcement learning algorithm to ensure the molecules produced align with new objectives. Although DrugImprover has demonstrated encouraging outcomes, its effectiveness is limited due to its dependence on the less complex LSTM network architecture, which might lead to limited scalability and capacity, contextual understanding.

On the other hand, generative pretrained language models have shown exceptional performance in diverse areas, from natural language understanding and generation exemplified by ChatGPT~\citep{wu2023brief, ouyang2022training}, to text-to-video conversion as seen in SORA~\citep{liu2024sora}, and in programming and code generation through platforms like PG-TD~\citep{zhang2023planning} and Copilot~\citep{nguyen2022empirical}. Nevertheless, in the field of drug discovery, tools such as DrugGPT~\citep{li2023druggpt}, ChatDrug~\citep{liu2024conversational} and ChemGPT~\citep{frey2023neural}, despite making some initial strides, have not yet achieved performance on par with their counterparts in other domains. The field of drug discovery remains in expectation of breakthroughs similar to the ones achieved by LMs in other domains.

Several challenges have hindered the impact of Language Models (LMs) on drug design: Firstly, the molecules created need to satisfy multiple criteria such as solubility and synthesizability, and must also secure a high docking score against a specific target site~\citep{liu2024binding}. However, existing drug discovery LMs generally only undergo pretraining with molecules and do not focus on enhancing multiple attributes simultaneously. Secondly, given that drugs sharing similar chemical structures tend to exhibit comparable biological or chemical effects~\citep{bender2004molecular}, it is crucial to optimize the drug while preserving the beneficial chemical structure of the original molecule. Lastly, the current methodologies in drug discovery LMs primarily focus on maximizing likelihood during the decoding phase, rather than customizing the optimization process to meet specific goals.

\fix{
REINVENT 4~\citep{he2021molecular, he2022transformer, loeffler2024reinvent} achieves the state of art performance in LM-based
drug optimization by using the Transformer model and conducting experiments with randomly selected ligand pairs that maintain constrained Tanimoto Similarity.
While this method successfully produces ligands with high Tanimoto Similarity, it encounters difficulties in consistently improving the properties of the original drug and tends to simply maximize the likelihood of molecule generation.
}

\subsection*{Generalizable Insights about Machine Learning in the Context of Healthcare}

In this work, we propose \algname, a novel scaffold-based GPT with three-stage optimization process for drug optimization. This optimization process is designed to enhance existing drugs to rapidly evolving virus variants and cancer cells, overcoming the limitation of earlier drug optimization efforts.
The contributions of \algname are summarized as follows:

$\bullet$ A novel framework for drug optimization using GPT, featuring a three-step optimization process, the first of its kind to the best of our knowledge. \fix{The underlying motivation is that each stage is complementary to each other, enhances the performance of the preceding one. Furthermore, we have conducted ablation studies to illustrate the performance gains achieved at each step.}

$\bullet$ A scaffolds-based \fix{GPT}  with a novel two-phase incremental pre-training specifically designed for drug optimization. The motivation for incremental training is to conduct local optimization before embarking on global optimization by breaking down the whole training corpus into knowledge pieces in an incremental order. We also conducted an ablation study in table 3 to demonstrate the effectiveness of the incremental training.

$\bullet$ A novel token-level decoding optimization strategy, $\TOPN$, utilizes pretrained \fix{GPT} to enable controlled, reward-guided generation that aligns with targeted objectives.

$\bullet$ Through extensive experiments and ablation studies on real-world viral and cancer-related benchmarks, we demonstrate that \algname outperforms the competing baselines
and reliably improves upon existing molecules/drugs in terms of desired targeted objectives while preserving original scaffold, resulting in superior drug candidates.

\section{Related Work}\label{sec:related}

\subsection{Reinforcement Learning and Finetuning}

Reinforcement learning~\citep{tan2022reinforcement, liu2023active, liu2023blending}  has become a fundamental strategy in drug design~\citep{born2021paccmannrl, guimaraes2017objective, neil2018exploring, olivecrona2017molecular, popova2018deep,  staahl2019deep, tan2022drlinker, wang2022reinforcement, zhang2023universal, zhou2019optimization}, focusing on optimizing rewards that aggregate predicted property scores from various pharmaceutical predictors. Traditional reinforcement learning methods in drug discovery often neglected molecular structure constraints, leading to significant structural changes and unsynthesizable compounds. Our research differs by refining existing drugs to enhance their attributes without redesigning them from scratch and using reinforcement learning to improve a pre-trained language model generator, rather than starting a new. We also incorporate methods like Advantage-aligned Policy Optimization (APO)~\citep{liu2023drugimprover}, which assigns rewards based on advantage preference over the original molecule, to fine-tune a Transformer model, ensuring it aligns with multiple pharmaceutical objectives while preserving molecular structure. This approach, which includes controllable decoding, refines the model beyond traditional reinforcement learning fine-tuning stages.

\subsection{Planning with GPT Models}

Several studies have leveraged planning algorithms to enhance text outputs for a variety of NLP tasks. These include approaches like beam search optimization, machine translation improvements~\citep{scialom2021beam,leblond2021machine,chaffin2021ppl}. The PG-TD~\citep{zhang2023planning} method is tailored for code generation using a singular reward function, whereas ERP~\citep{liu2024erp} introduces a novel concept by considering the certainty of each generated token along with an \emph{e}-step forward entropy measurement to gauge potential outcomes. It has been shown that ERP effectively balances exploration and exploitation within molecular structures, leading to the discovery of high-reward molecules. Unlike previous studies that focus solely on planning with pre-trained language models, our approach incorporates a novel decoding optimization as a critical final step in the algorithm. Moreover, our focus is on optimizing an existing drug rather than creating a De Novo one from the ground up.

\subsection{Drug Optimization}

Recent drug optimization efforts have focused primarily on a limited array of drug properties while often disregarding the docking score, an essential metric for evaluating structural compatibility with a target~\citep{zhou2019optimization, erikawa2021mermaid}. DrugEx v3~\citep{liu2023drugex} seeks to resolve this deficiency by using 3D molecular graphs that encapsulate more comprehensive data such as chemical valence rules. Nevertheless, this method's complexity poses challenges in generating molecules that closely resemble the originals. The resulting lower similarity could account for the diminished efficacy of graph-based methods, as deviations from the original molecular structures result in the loss of vital chemical properties. Conversely, our approach manages to preserve a decent level of similarity. 

Diffusion-based efforts~\citep{alakhdar2024diffusion,morehead2024geometry} either focus on De Novo drug discovery while neglecting the essential similarity to the original molecule, or overlook the docking score towards a binding target.
Molsearch \citep{sun2022molsearch} is a Monte Carlo tree search (MCTS)-driven approach for molecular generation and optimization, 
and MIMOSA \citep{fu2021mimosa} is a GNN sampling-based method leveraging graph-based molecular optimization.
DrugImprover~\citep{liu2023drugimprover} effectively begins to redefine the drug optimization challenge by employing reinforcement learning with a mix of multiple objectives as rewards. Still, it employs an LSTM in its generative model, which faces limitations in scalability, capacity, and understanding context. In contrast, our method applies a Transformer as the primary generative model, enhanced further with Advantage-aligned Policy Optimization (APO) for exploratory purposes and an optimized decoder for superior performance.

\subsection{Language Models for Drug Discovery}

Large language models like MolGPT~\citep{bagal2021molgpt} and ChemGPT~\citep{frey2023neural} have been utilized in molecule generation and drug discovery~\citep{bagal2021molgpt,rothchild2021c5t5,wang2022transformer}, using formats like SMILES~\citep{weininger1988smiles} and SELFIES~\citep{krenn2020self} to standardize molecular representation. These models have shown promise in drug design, outperforming traditional methods in some predictive capacities. However, as noted by ~\citet{murakumo2023llm}, using pre-trained language models typically results in minor molecular modifications and serves primarily supportive roles in the design process. We adopt the SMILES format and combine the strengths of GPTs with reinforcement learning. Unlike previous efforts that relied solely on LLMs, our approach uses RL fine-tuning to align with multiple pharmaceutical objectives and employs controlled decoding to guide the GPT model in generating more effective molecular structures.

Further developments include transformer-based models like REINVENT 4 ~\citep{he2021molecular, he2022transformer, loeffler2024reinvent}, which mostly focuses on pretraining. While pretraining helps in generating molecules similar to those in the training dataset, it also inherently restricts the exploration scope due to biases in the training data.
REINVENT4 uses the original molecule as input, generates molecules that are very similar to the original ones. This likely contributes to its relatively poor performance in other metrics, as the generated molecules exhibit minimal changes in chemical properties. 
In contrast, our proposed method, which uses SMILES and scaffolds as prompts, achieves a well-balanced trade-off between diversity and similarity in the generated and original molecules, potentially leading to improved performance.

\section{Preliminaries}

In the following sections, we detail MDP, \fix{Language Model} (LM) and drug discovery, complete with their mathematical notations, and integrate them within the framework of Markov decision processes.

\textbf{Markov decision processes.}
Let us define a finite-horizon Markov decision process (MDP)~\citep{puterman2014markov} $\MDP_0 = \langle \stateSpace, \actionSpace,\horizon,\transDynamics,\RFunc \rangle$.
In this context, $\stateSpace$ represents a finite set of states, while $\actionSpace$ comprises a finite set of actions. The term $\horizon$ denotes the planning horizon. The function $\transDynamics$, defined as $\transDynamics: \stateSpace \times \actionSpace \rightarrow \stateSpace'$, describes the deterministic transition dynamics that combine a state $\state$ with an action $\action$, with an episode concluding once the agent executes the termination action. Additionally, the reward function $\RFunc: \stateSpace \times \actionSpace \rightarrow \mathbb{R}$ assigns scores exclusively to complete molecules, assigning a reward of 0\SJ{depends on results / which run/hyperparam to use, will check on weekends} to partial molecules.
The effectiveness of a policy can be evaluated using the $\QFunc$-value function, denoted as $\QFunc^{\policy}: \stateSpace \times \actionSpace \rightarrow \mathbb{R}$, and defined by the following equation:
$\QFunc^{\policy}(\state, \action) := \mathbb{E}^{\policy}\left[\sum_{t=0}^{\horizon}\RFunc(\state_t, \action_t) \mid \state_0 = \state, \action_0 = \action\right],$
where the expectation is based on the trajectory determined by the policy $\policy$. The associated value function is given by:
$\VFunc^{\policy}(\state) := \mathbb{E}_{\action \sim \policy(\cdot|\state)}\left[\QFunc^{\policy}(\state, \action)\right]$.

\textbf{LM.} We define the state space $\stateSpace$ as the set of all possible molecule, where each molecule is represented as a state $\state$ that includes a start token $\bracket{\text{BOS}}$, a molecule with SMILES~\citep{weininger1988smiles} representation string, and a termination action $\bracket{\text{EOS}}$. 
We define the set of complete molecules as
\begin{equation}
    \hypothesis_{\horizon} := \curlybracket{\text{[BOS]} \circ \mathbf{v} \circ \text{[EOS]}~|~\mathbf{v}\in \mathcal{V}^*},
\end{equation}
where $\hypothesis_t \subseteq \stateSpace_t|_{t\in \bracket{\horizon}}$ represents the hypothesis space at step $t$ (sequence length $t$), $\mathcal{V}^*$ represents the Kleene closure of Transformer's vocabulary $\mathcal{V}$, with  $\mathcal{V}:=\actionSpace$, and $\circ$ indicating string concatenation. 
Each action $\action \in \actionSpace$ is represented as token $y \in \mathcal{V}$.
In this work, we train a GPT policy $\policy_{\theta}$ to acquire prior knowledge for generating valid molecules based on a given set of molecules $\Buffer$.
We define the generator policy $\policy_{\theta}$, with learned weights $\theta$, as the product of probability distributions: $\policy_{\theta}\paren{\mathbf{y}|\mathbf{x}}=\prod_{t=1}^{|\mathbf{y}|} \policy_{\theta}\paren{y_t|\mathbf{x},\mathbf{y}_{<t}}$, 
where  $\policy_{\theta}\paren{\cdot|\mathbf{x},\mathbf{y}_{<t}}$ is a distribution,  $\mathbf{x}$ is an input sequence, and $\mathbf{y}_{<1}=y_0:=\text{[BOS]}$.
The decoding process in text generation involves identifying the most likely hypothesis by optimizing the objective:
$\mathbf{y}^{\star}=\argmax_{\mathbf{y}\in \hypothesis_{\horizon}} \log \policy_{\theta}\paren{\mathbf{y}|\mathbf{x}}.$

\begin{figure*}[t]
        \centering
        \includegraphics[%
        width=15.5cm, 
        clip={0,0,0,0}]{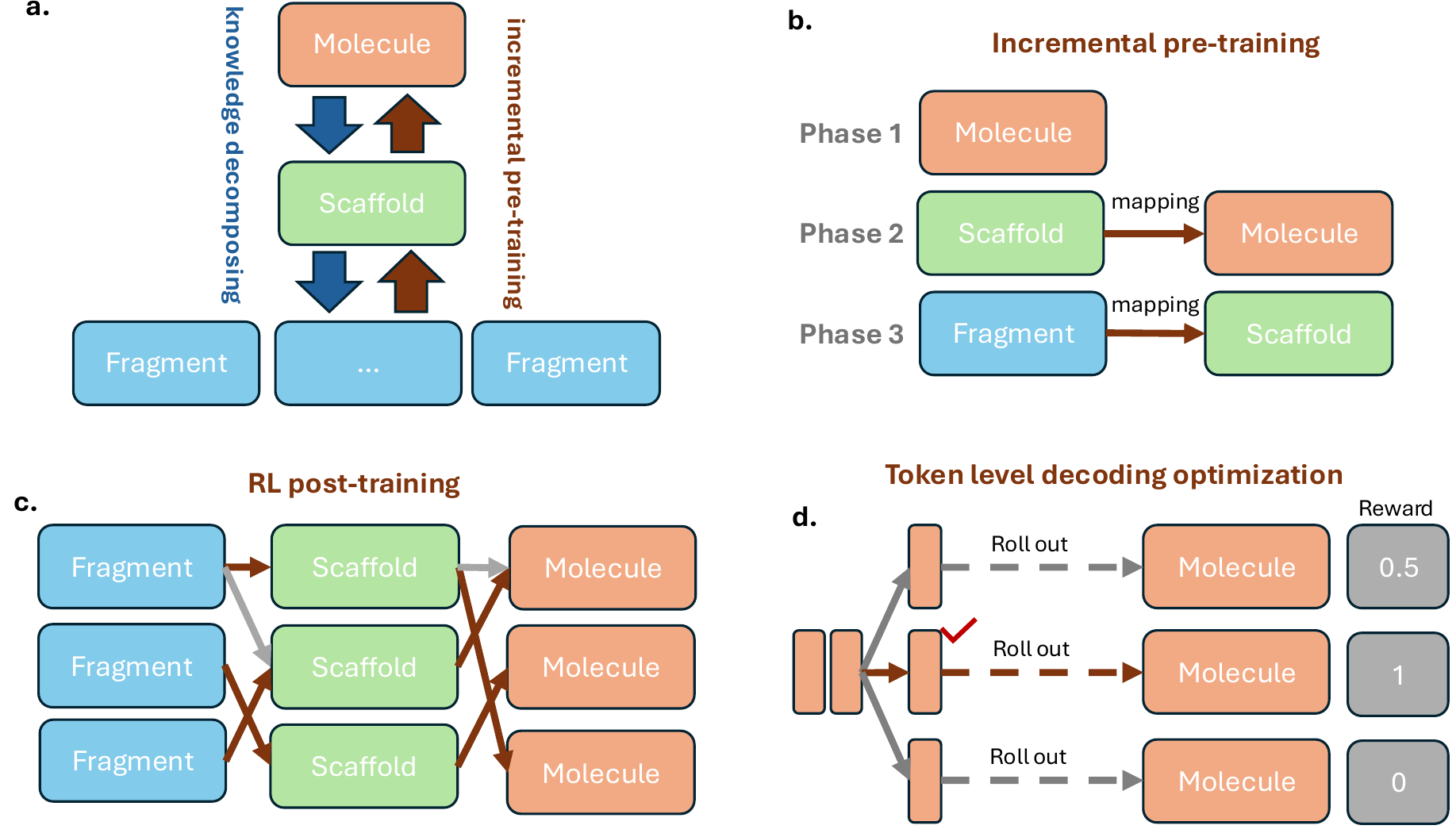}
    \caption{Overview of \algname. (a) \algname decomposes molecular knowledge top-down into fragments and then reconstructs it bottom-up through linking. (b) In Stage 1, it proposes an incremental pretraining strategy. (c) In Stage 2, it undergoes reinforcement learning-based post-training. (d) In Stage 3, it introduces a token-level controllable decoding optimization approach. \algname is the first to introduce a three-stage optimization framework, where each stage complements the others.
    }
  \label{fig:scaffoldgpt_overview}
\end{figure*}

\textbf{Drug Optimization.} Given an initial drug candidate $X=\paren{x_1,\cdots,x_{\horizon}}$ and a drug optimization policy $\policy_{\theta}$, the goal in drug optimization is to find the optimal policy $\policy_{\theta^*}$ that
maximize the following objective:
\begin{equation}\label{eq:drug_generation_objective}
\policy_{\theta^*}=
\argmax_{\policy_{\theta}}
\expctover{X \sim {\stateDist_0}}{\RFunc(Y)-\RFunc\paren{X}|\theta,X}, 
\end{equation}
where 
 $Y=\policy_{\theta}\paren{\cdot\given X}, Y_{1:\horizon}=\paren{y_1,\ldots,y_t,\ldots,y_{\horizon}},y_t\in \mathcal{V}.$

\textbf{Limitation of previous works:} 
DrugImprover, which utilizes LSTM networks, has limitations in scalability, capacity, and contextual understanding, especially when compared to versions that use GPT Models. 
The current state of the art, REINVENT 4, employs the Transformer architecture; however, {it mainly focuses on pretraining with constrained similarity, which restricts its capability to explore molecular spaces that might offer high rewards beyond its training set.}
In this work, we address these limitations by proposing \algname.

\section{The \algname Algorithm}

\begin{algorithm}[t]
    \caption{\algname}\label{alg:scaffoldgpt}
    \begin{algorithmic}[1] 
    \Require { GPT-based generator policy $\Gen$; {critics $\mathbf{\critic}$}.}
    \State Initialize $\Gen$ with GPT2-like Transformer with random weight $\theta$.
    \LineComment{/* Stage 1: Pre-training GPT-based generator */}
    \State Build the training corpus $\corpus_\text{Phase 1}$ (\ref{eq:phase1_corpus}), $\corpus_\text{Phase 2}$ (\ref{eq:phase2_corpus}).
    \State Pre-train BPE tokenizer and $\Gen$ on  $\corpus_\text{Phase 1}$ via CLM objective~\eqref{eq:pretrain-loss}.
    \State Pre-train $\Gen$ on  $\corpus_\text{Phase 2}$.

    \LineComment{/* Stage 2: APO fine-tuning */}

    \For{$n=1,\ldots,\ENum$}
            \State $\state_0 \sim \rho_0$, { \text{where }$\rho_0\in \Delta\paren{\RationaleBuffer}$}.
            \State Generate $Y_{1:\horizon}=\paren{y_t,\ldots,y_{\horizon}}\sim \Gen\paren{\cdot|S}$.
            \State Compute advantage preference
            {$\RFunc^\text{AP}$}
            by \eqref{eq:reward}\eqref{eq:advantage_preference}.
            \State Update generator $\theta$
            via policy gradient by \eqref{eq:adv:preference}.
    \EndFor
    
    \LineComment{/* Stage 3: Token-level Decoding Optimization */}
    \State Optimize the generation of $\Gen$ via $\TOPN$ \eqref{eq:topN} decoding strategy.    
    \end{algorithmic}
\end{algorithm}

\textbf{Stage 1. Pretrain a GPT generator.} 
Let us note the GPT generator policy as $\policy_{\theta}$, which computes the probability $p$ of the occurrence of the $t^{th}$ token in a target molecule $Y$. It takes into account all preceding tokens $\mathbf{y}_{<t} = \bracket{y_1, ..., y_{t-1}}$ in the target, as well as the scaffold compound $S$, which is noted as
$\policy_{\theta}\paren{y_t \given \mathbf{y}_{<t},S}=p\paren{y_{t}|\mathbf{y}_{<t},S}.$
The parameters $\theta$ of the generator policy $\policy_{\theta}$ are trained using the training corpus set through the minimization of the negative log-likelihood (NLL) for the complete SMILES strings across the entire set.
This process is described as follows:
\begin{align}\label{eq:pretrain-loss}
    NLL =  - \log \Prob\paren{Y|S} 
    =-\sum_{t=1}^{\horizon} \log \Prob\paren{y_{t}|y_{t-1},...,y_{1},S}
   =-\sum_{t=1}^{\horizon} \log \policy_{\theta}\paren{y_t|y_{1:t-1}, S},
\end{align}
where $\horizon$ signifies the total number of tokens related to $Y$. The NLL quantifies the probability of converting a specific scaffold into a designated target molecule. 

In this project, we employ pre-training to harness large quantities of unlabeled text to construct a basic foundation model of language understanding. This foundation model can subsequently be fine-tuned and tailored to meet various specialized goals.
In this work, we propose a novel framework for pre-training a Transformer and linking scaffolds with complete molecules, based on a SMILES (Simplified Molecular Input Line Entry System) \citep{weininger1988smiles}  string representation of the molecule. Here, we define $\mathcal{S}$ as the scaffold space and $\mathcal{Y}$ as the target molecule space. With $\Pair=\curlybracket{\paren{s,y}|s,y\in \mathcal{S} \times \mathcal{Y}}$, we denote the set of scaffold and molecular pairs from $\mathcal{S}$ and $\mathcal{Y}$. In this notation, $s$ represents the scaffold of the target molecule, and $y$ is the corresponding target molecule.
We initially pre-trained our tokenizer using the Byte Pair Encoding (BPE) method \fix{\citep{gage1994new_bpe, sennrich2015neural_bpe}}.  
Building on the pre-trained BPE tokenizer, we propose a two-phase incremental training approach,
as illustrated in \figref{fig:drugimprover_framework}, to notably enhance the model's ability to improve the validity of inferring the target molecule from its scaffold.

\paragraph{Incremental \fix{pre-}training.} 
The rationale for incremental pre-training is to conduct local optimization before embarking on global optimization. Therefore, we divide our training into two phases.
In the first phase, we focus on training a GPT exclusively for molecules using Causal Language Modeling (CLM). CLM utilizes an autoregressive method where the model is trained to predict the next token in a sequence by considering only the tokens that precede it. The phase 1 corpus is designed as follows:
\begin{figure*}[t]
        \centering
        \includegraphics[%
        width=13cm, 
        clip={0,0,0,0}]{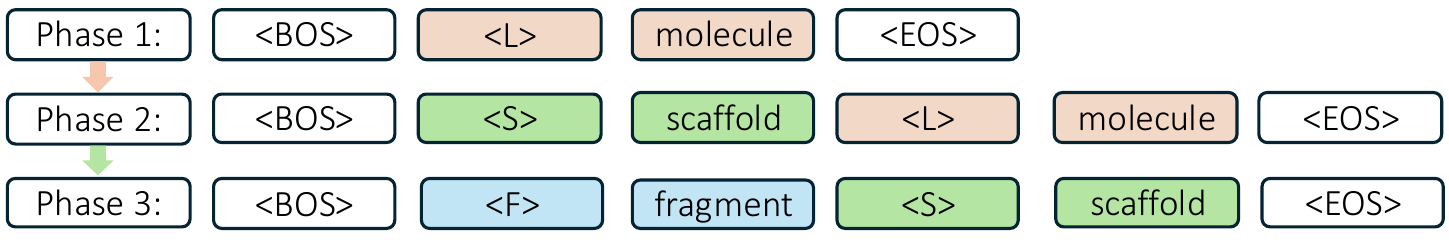}
    \caption{
    Two-phase incremental training of \algname. The first phase concentrates on recognizing the molecule, while the second phase builds connections between its scaffold and the original molecule. \fix{This process can be extended into a three-phase pretraining framework by establishing connections among the molecule’s fragments, scaffold, and original molecule.}
    }
  \label{fig:drugimprover_framework}
\end{figure*}
\begin{equation}\label{eq:phase1_corpus}
\corpus_{\text{Phase 1}}=\curlybracket{[BOS],\tuple{L},\underbrace{y_1,\cdots,y_{\horizon}}_{\text{target molecule Y}},[EOS]}, \fix{\tuple{L} \text{is the token for ligand.}}
\end{equation}
In the second phase, building upon the success of the GPT model developed in phase 1, which demonstrated high accuracy in molecular generation, we advance the training by focusing on pairs of scaffolds and molecules using CLM. The phase 2 corpus is as follows:
\begin{align}\label{eq:phase2_corpus}
\corpus_{\text{Phase 2}}=\curlybracket{[BOS],\tuple{S},\underbrace{s_1,\cdots,s_{\horizon}}_{\text{source scaffold S}},\tuple{L},\underbrace{y_1,\cdots,y_{\horizon}}_{\text{target molecule Y}},[EOS]}, \fix{\tuple{S} \text{: scaffold.}}
\end{align}
Consequently, the model can generate the appropriate molecule when given a scaffold as prompt. 
\fix{While this work primarily focuses on scaffolds, the proposed method can be  extended to fragments as well. A scaffold represents a complete core structure, while fragments may constitute parts of the scaffold or side chains attached to it. This approach could be further developed into a fragment-based third-phase pre-training strategy, with the phase 3 corpus outlined as follows:}
\begin{align}\label{eq:phase3_corpus}
\corpus_{\text{Phase 3}}=\curlybracket{[BOS],\tuple{F},\underbrace{f_1,\cdots,f_{\horizon}}_{\text{source fragment F}},\tuple{S},\underbrace{s_1,\cdots,s_{\horizon}}_{\text{source scaffold S}},[EOS]}, \fix{\tuple{F} \text{: fragment.}}
\end{align}
\fix{As a result, the model can produce a suitable scaffold when provided with a fragment as a prompt, and using the scaffold as the input prompt will generate a complete molecule.}
However, since a single scaffold can relate to multiple molecules and a fragment can correspond to multiple scaffolds, we further optimize the GPT-based generator policy, $\policy_{\theta}$, to focus on specific outcomes by implementing reinforcement learning fine-tuning in the subsequent stage.

\textbf{Stage 2. RL post-training/finetuning.} 
Fine-tuning a generative model is crucial for producing outcomes that meet specific objectives. In this study, we use the Advantage-alignment Policy Optimization (APO)~\citep{liu2023drugimprover} algorithm to fine-tune the pretrained GPT-based generator policy $\policy_{\theta}$. This approach steers the model from a given scaffold towards the targeted molecule, simultaneously improving multiple properties.

In this work, we adopt the define of reward function from \citet{liu2024erp}, and regarded each pharmaceutical property as a critic $C$ and got an ensemble critics $\mathbf{C}$ as follows:
\begin{align*}
[\critic^{\text{Druglikeness}},
\critic^{\text{Solubility}},
\critic^{\text{Synthesizability}}, 
{\critic^{\text{Docking}}},\critic^{\text{{Tanimoto}}}
],
\end{align*}
where each critic $\critic: Y \rightarrow \mathbb{R}$ {acts as a distinct evaluator for a specific pharmaceutical attribute}.
We built the reward function as follow:
\begin{align}\label{eq:reward1}
{\RFunc^{\text{norm}}\paren{Y}}:=
{\RFunc^{\text{norm}}\paren{
Y|S}} =\lambda\cdot\text{Norm}\paren{\critic^{\text{Tanimoto}}\paren{S,
Y}}
+
\sum_{i=0}^{|\mathbf{C}|-1}\lambda\cdot\text{Norm}\paren{\critic_i{\paren{Y}}}
\end{align}
where Norm is employed to standardize diverse attributes to a consistent scale. In this instance, Norm refers to the process of min-max normalization, which is used to adjust the attributes so they fit within the $\bracket{0,1}$ range. \fix{ We defer the details of critics to \secref{exp_config}.}

In this work, we use $\BON$~\citep{gao2023scaling} (Best of N) search to estimate the reward for a prompt (partial molecule). $\BON$ can be formulated as the following:
\begin{align}\label{eq:bon}
{\BON}&\paren{\mathbf{y}_{<i},N,\RFunc}|_{S,p,k}= \max_{\mathbf{Y}_{j}\in \curlybracket{\mathbf{Y}_1,\cdots,\mathbf{Y}_{N}}} \RFunc\paren{ \mathbf{Y}_{j}},\\
\text{where } Y_j=&\bracket{\mathbf{y}_{<i},y_i,\cdots,y_{\horizon}}_j, \text{and }
y_i \sim \TOPPK\paren{\mathbf{y}_{<i},p, k}|_{S}.\notag
\end{align}

\noindent where \TOPPK~\citep{liu2024erp} is defined as follows:
\begin{align}\label{eq:toppk}
    &\text{\TOPPK}\paren{\mathbf{y}_{<i}, p,k}|_{S}= \actionSpace_{\mathbf{y}_{<i}}, 
   \text{where~} \actionSpace_{\mathbf{y}_{<i}}=\curlybracket{y_1,\ldots,y_{j}}, y_i\in \mathcal{V},  \\ 
    & j= \min \curlybracket{\argmin_{j'}\sum_{i=1}^{j'}\policy_{\theta}\paren{y_i|S,\mathbf{y}_{<i}} \geq p, k},
    \text{and } \policy_{\theta}\paren{y_g|S,\mathbf{y}_{<i}}>\policy_{\theta}\paren{y_h|S,\mathbf{y}_{<i}}, \text{if } g<h, \notag
\end{align}
where $p\in (0,1]$ represents the maximum cumulative probability, and $k$ denotes the maximum number of candidates for the next tokens.

For each pair consisting of a scaffold and a molecule, we create 8 new molecules from the scaffold using \TOPPK~\eqref{eq:toppk}  and \BON~\eqref{eq:bon} method to select the best one (the one with the highest normalized reward) to serve as the foundation for the final reward calculation.
\begin{align}\label{eq:reward}
{\RFunc_c\paren{Y|S}}= \RFunc^{\text{norm}}{\paren{Y}}|_{S}, ~~{Y} \in \BON\paren{y_0,N,\RFunc}|_{S,p,k}.
\end{align}
APO makes policy gradient based on the advantage preference~\citep{liu2023drugimprover}, which is defined as
\begin{equation}\label{eq:advantage_preference}
\RFunc^\text{AP}\paren{Y_{1:\horizon},S} ={\RFunc_c\paren{Y_{1:\horizon}} - \RFunc_c\paren{S}},
\end{equation}
and perform APO policy gradient with follows:
\begin{align}\label{eq:adv:preference}
     \gradient=& \textstyle \mathbb{E}_{S\sim {\rho_0},{Y_{1:\horizon}\sim \Gen\paren{\cdot|S}} } 
     \left[ \nabla_{\theta}\log \Gen\paren{Y_{1:\horizon}|S}
     \cdot \RFunc^\AP\paren{S,Y_{1:\horizon}}\right],
\end{align}
\paragraph{Stage 3. Token-level Controllable decoding generation.}\label{decoding_strategy}
Ultimately, the current GPT-based decoder focuses mainly on maximizing likelihood, neglecting specific metrics of interest. This approach limits its effectiveness in optimizing objectives that diverge from those in its training set, particularly in generating desired molecules. In this study, we introduce controllable decoding after fine-tuning with APO. We present a new approach, \TOPN, to direct the generation process towards enhancements in the optimization objective, as detailed below:
\begin{align}\label{eq:topN}
Y^{\star}&\sim\curlybracket{[\mathbf{y}_{<i},y_i,\cdots,y_{\horizon}]}, 
\text{~where~} y_i\sim \text{\TOPN}\paren{\mathbf{y}_{<i}, p,k, n}|_{S}  \\ \notag
    &\text{~\TOPN}\paren{\mathbf{y}_{<i}, p,k, n}|_{S}= \actionSpace_{\mathbf{y}_{<i}}, 
    \text{~} \actionSpace_{\mathbf{y}_{<i}}=\curlybracket{y_1,\ldots,y_{n}}, y_i\in \mathcal{V}, |\actionSpace|\leq k, \notag \\ \notag
    &\text{~and } 
    \RFunc\paren{\BON\paren{\mathbf{y}_{<i}\circ y_g,N,R}|_{S,p,k}}
    \geq\RFunc\paren{\BON\paren{\mathbf{y}_{<i}\circ y_h,N,R}|_{S,p,k}}, \forall g<h, \notag
\end{align}
where $n\leq k$ denotes as top $N$ candidate of next tokens with regard to $\BON$ function.

\begin{remark}
$\TOPN$ differs from $\TOPP$, $\TOPK$, and $\TOPPK$ in that it is measured based on maximum reward, whereas the others are measured based on maximum likelihood. $\TOPN$ is also distinct from $\BON$ in that $\BON$ is optimized at the sequence level, while $\TOPN$ is optimized at the token level.
\end{remark}

\section{EXPERIMENTS}\label{experiments}

\subsection{Experimental configuration}\label{exp_config}

{\textbf{The language model.} We use GPT-2-like Transformers for causal language modeling, employing the standard 11M Drug-like Zinc dataset for training. Entries with empty scaffold SMILES are excluded, and we adopt a 90/10 split for training and validation, respectively. The training process is structured into three phases: pretraining, fine-tuning, and decoding optimization, as outlined in Algorithm \algname (See appendix for more details).}

\textbf{Baselines.}
{
We compare against baseline models: DrugImprover~\citep{liu2023drugimprover}, which utilizes an LSTM-based generator with APO fine-tuning; 
Molsearch \citep{sun2022molsearch}, a Monte Carlo tree search (MCTS)-driven approach for molecular generation and optimization; MIMOSA \citep{fu2021mimosa}, a sampling-based method leveraging graph-based molecular optimization; and \fix{DrugEx v3 \citep{liu2023drugex}, which leverages transformer-based reinforcement learning for scaffold-guided drug optimization}. Additionally, we incorporate the model proposed by \citet{he2021molecular, he2022transformer,  loeffler2024reinvent}, which trains a transformer to adhere to the Matched Molecular Pair (MMP) guidelines~\citep{kenny2005structure,tyrchan2017matched}.
Specifically, given a set $\{\{X,Y,Z\}\}$, where $X$ represents the source molecule, $Y$ denotes the target molecule, and $Z$ signifies the property change between $X$ and $Y$, the model learns a mapping from $\{X, Z\} \in \ensuremath{\mathcal{X}} \times \ensuremath{\mathcal{Z}} \implies Y \in \ensuremath{\mathcal{Y}}$ during training. Here, $\ensuremath{\mathcal{X}} \times \ensuremath{\mathcal{Z}}$ denotes the input space, and $\ensuremath{\mathcal{Y}}$ denotes the target space. They defined six different types of property changes for $Z$, including MMP for user-specified alterations, various similarity thresholds, and scaffold-based modifications where molecules share the same scaffold or a generic scaffold. More specifically,
\begin{itemize}[noitemsep, topsep=0pt, partopsep=0pt]
    \item MMP: there exists user-specified property changes between molecule $X$ and $Y$.
    \item Similarity $\geq 0.5$: tanimoto similarity between molecule $X$ and $Y$ is larger than 0.5.
    \item Similarity $\in[0.5, 0.7)$: the tanimoto similarity of pair $\paren{X,Y}$ is between 0.5 and 0.7.
    \item Similarity $\geq 0.7$: tanimoto similarity between molecule $X$ and $Y$ is larger than 0.7.
    \item Scaffold: molecule $X$ and $Y$ share same scaffold.
    \item Scaffold generic: molecule $X$ and $Y$ share same generic scaffold.\XL{cite}
\end{itemize}
}
\fix{All baseline models are fine-tuned on the cancer and COVID datasets following their respective fine-tuning methods.}

\textbf{Dataset.}
We employ, from the most recent Cancer and COVID dataset of \citet{liu2023drugimprover}, 1 million compounds from the ZINC15 dataset docked to the 3CLPro~(PDB ID: 7BQY) protein associated with SARS-CoV-2 and the RTCB (PDB ID: 4DWQ) human cancer protein.

{
\textbf{Critics and evaluation metric.} 
In this study, we evaluate the efficacy of \algname in generating molecules with desirable attributes within the context of pharmaceutical drug discovery.
{We leverage the RDKit  \citep{landrum2016rdkit} chemoinformatics package and employ various performance metrics as follows:}
{\textbf{Validity} measures if the generated SMILES is valid in syntax.}
\textbf{Druglikeness} measures the likelihood of a molecule being a suitable candidate for drug development.
\textbf{Solubility} assesses the likelihood of a molecule's ability to mix with water, commonly referred to as the water-octanol partition coefficient (LogP). 
\textbf{Synthetizability} quantifies the ease (score of 1) or difficulty (score of 10) associated with synthesizing a given molecule \citep{ertl2009estimation}.
\textbf{Docking Score} assesses the drug's potential to bind and inhibit the target site. 
To enable efficient computation, we employ a docking surrogate model (See \appref{app:surrogate_model}) to output this score.
{
\textbf{Similarity:} We use Tanimoto similarity to evaluate the similarity between original SMILES and generated SMILES.
\textbf{Average Top 10\% Norm Reward} is the average of the normalized reward of the top 10\% of molecules based on their average normalized reward.
\textbf{Average Norm Reward} is the average of the normalized values of the docking score, druglikeness, synthesizability, solubility, and similarity across all valid molecules. This is the most important metric. \fix{ Evaluations are based on a sample of 1,280 molecules.}
}
}
\begin{remark}
A similarity score that is too high results in low structural diversity, while a score that is too low suggests the molecules have drifted too far from the original. Neither extreme is desirable. Our goal is to achieve a balanced level of similarity with meaningful variation. In this work, the primary optimization objective is the average normalized reward.
\end{remark}

\begin{table*}[t!]
\setlength{\tabcolsep}{4pt}
   \centering
    {\small
    \scalebox{0.48}{
    \begin{tabular}{l l c c c c c c c c c c }
        \toprule
        \textbf{Target} %
        & \textbf{Algorithm}
        & {\makecell[c]{Validity~$\uparrow$}}
        & {\makecell[c]{Avg \\ Norm Reward~$\uparrow$$^{{\star}}$}}
        & {\makecell[c]{Avg Top 10 \% \\ Norm Reward~$\uparrow$}}
        & {\makecell[c]{Docking ~$\downarrow$}}
        & {\makecell[c]{Druglikeliness ~$\uparrow$}}
        & {\makecell[c]{Synthesizability ~$\downarrow$}}
        & {\makecell[c]{Solubility ~$\uparrow$}}
        & {\makecell[c]{Similarity~$\uparrow$}}
        
        \\
        \midrule
        \makecell[l]{\textbf{3CLPro}} %
        &  \textbf{\makecell[l]{Original}}
        &  \makecell[l]{-}
        &  \makecell[l]{0.533}
        &  \makecell[l]{{0.689} }
        &  \makecell[l]{-8.698 }
        &  \makecell[l]{0.682 }
        &  \makecell[l]{3.920 }
        &  \makecell[l]{2.471 }
        &  \makecell[l]{-}
        \\
        (PDBID:
        &  \textbf{\makecell[l]{MMP \citep{loeffler2024reinvent}}}
        &  \makecell[l]{0.995 $\pm$ 0.001}
        &  \makecell[l]{0.628 $\pm$ 0.001}
        &  \makecell[l]{0.718 $\pm$ 0.000}
        &  \makecell[l]{-8.259 $\pm$ 0.004}
        &  \makecell[l]{0.691 $\pm$ 0.001}
        &  \makecell[l]{2.682 $\pm$ 0.004}
        &  \makecell[l]{3.109 $\pm$ 0.020}
        &  \makecell[l]{0.862 $\pm$ 0.000}
        \\
       \ 7BQY)
        &  \textbf{\makecell[l]{Similarity ($\geq$ 0.5) \citep{loeffler2024reinvent}}}
        &  \makecell[l]{0.995 $\pm$ 0.001}
        &  \makecell[l]{0.615 $\pm$ 0.000}
        &  \makecell[l]{0.706 $\pm$ 0.001}
        &  \makecell[l]{-8.165 $\pm$ 0.024}
        &  \makecell[l]{0.697 $\pm$ 0.004}
        &  \makecell[l]{2.621 $\pm$ 0.006}
        &  \makecell[l]{3.180 $\pm$ 0.029}
        &  \makecell[l]{0.782 $\pm$ 0.001}
        \\
        \textbf{ }
        &  \textbf{\makecell[l]{Similarity ([0.5, 0.7)]) \citep{loeffler2024reinvent}}}
        &  \makecell[l]{0.995 $\pm$ 0.001}
        &  \makecell[l]{0.612 $\pm$ 0.001}
        &  \makecell[l]{0.701 $\pm$ 0.001}
        &  \makecell[l]{-8.187 $\pm$ 0.010}
        &  \makecell[l]{0.691 $\pm$ 0.001}
        &  \makecell[l]{2.611 $\pm$ 0.009}
        &  \makecell[l]{3.240 $\pm$ 0.014}
        &  \makecell[l]{0.756 $\pm$ 0.003}
        \\
        \textbf{ }
        &  \textbf{\makecell[l]{Similarity ($\geq$ 0.7) \citep{loeffler2024reinvent}}}
        &  \makecell[l]{0.995 $\pm$ 0.001}
        &  \makecell[l]{0.628 $\pm$ 0.001}
        &  \makecell[l]{0.718 $\pm$ 0.001}
        &  \makecell[l]{-8.214 $\pm$ 0.002}
        &  \makecell[l]{0.691 $\pm$ 0.002}
        &  \makecell[l]{2.717 $\pm$ 0.002}
        &  \makecell[l]{3.080 $\pm$ 0.016}
        &  \makecell[l]{0.881 $\pm$ 0.002}
        \\
        \textbf{ }
        &  \textbf{\makecell[l]{Scaffold \citep{loeffler2024reinvent}}}
        &  \makecell[l]{0.995 $\pm$ 0.001}
        &  \makecell[l]{0.602 $\pm$ 0.001}
        &  \makecell[l]{0.703 $\pm$ 0.002}
        &  \makecell[l]{-8.116 $\pm$ 0.002}
        &  \makecell[l]{0.695 $\pm$ 0.001}
        &  \makecell[l]{2.728 $\pm$ 0.008}
        &  \makecell[l]{2.968 $\pm$ 0.038}
        &  \makecell[l]{0.776 $\pm$ 0.001}
        \\
        \textbf{ }
        &  \textbf{\makecell[l]{Scaffold Generic \citep{loeffler2024reinvent}}}
        &  \makecell[l]{0.994 $\pm$ 0.001}
        &  \makecell[l]{0.617 $\pm$ 0.001}
        &  \makecell[l]{0.710 $\pm$ 0.002}
        &  \makecell[l]{-8.179 $\pm$ 0.012}
        &  \makecell[l]{0.701 $\pm$ 0.000}
        &  \makecell[l]{2.645 $\pm$ 0.008}
        &  \makecell[l]{3.090 $\pm$ 0.029}
        &  \makecell[l]{0.801 $\pm$ 0.000}
        \\
        \textbf{ }
        &  \textbf{\makecell[l]{DrugImprover \citep{liu2023drugimprover}}}
        &  \makecell[l]{0.884 $\pm$ 0.005}
        &  \makecell[l]{0.432 $\pm$ 0.002} 
        &  \makecell[l]{0.493 $\pm$ 0.005} 
        &  \makecell[l]{-6.726 $\pm$ 0.007}
        &  \makecell[l]{0.506 $\pm$ 0.002}
        &  \makecell[l]{\textbf{1.306} $\pm$ 0.010}
        &  \makecell[l]{2.057 $\pm$ 0.011}
        &  \makecell[l]{0.531 $\pm$ 0.002}
        \\
        \textbf{ }
        &  \textbf{\makecell[l]{Molsearch \citep{sun2022molsearch}}}
        &  \makecell[l]{\textbf{1.000} $\pm$ 0.001}
        &  \makecell[l]{0.616 $\pm$ 0.001} 
        &  \makecell[l]{0.726 $\pm$ 0.002} 
        &  \makecell[l]{-8.855 $\pm$ 0.040}
        &  \makecell[l]{0.686 $\pm$ 0.001}
        &  \makecell[l]{3.105 $\pm$ 0.006}
        &  \makecell[l]{2.452 $\pm$ 0.008}
        &  \makecell[l]{\textbf{0.969} $\pm$ 0.001}
        \\
        \textbf{ }
        &  \textbf{\makecell[l]{MIMOSA \citep{fu2021mimosa}}}
        &  \makecell[l]{0.985 $\pm$ 0.008}
        &  \makecell[l]{0.622 $\pm$ 0.001} 
        &  \makecell[l]{0.734 $\pm$ 0.002} 
        &  \makecell[l]{-8.800 $\pm$ 0.015}
        &  \makecell[l]{0.677 $\pm$ 0.004}
        &  \makecell[l]{3.105 $\pm$ 0.008}
        &  \makecell[l]{2.711 $\pm$ 0.010}
        &  \makecell[l]{\underline{0.959} $\pm$ 0.001}
        \\
        \textbf{ }
        &  \textbf{\makecell[l]{\fix{DrugEx v3} \citep{liu2023drugex}}}
        &  \makecell[l]{\textbf{1.000} $\pm$ 0.001}
        &  \makecell[l]{0.524 $\pm$ 0.001} 
        &  \makecell[l]{0.613 $\pm$ 0.001} 
        &  \makecell[l]{-8.089 $\pm$ 0.013}
        &  \makecell[l]{0.583 $\pm$ 0.002}
        &  \makecell[l]{3.095 $\pm$ 0.005}
        &  \makecell[l]{3.932 $\pm$ 0.008}
        &  \makecell[l]{{0.495} $\pm$ 0.001}
        \\
        \textbf{ }
        &  \textbf{\makecell[l]{\algname (w/o APO \& \TOPN )}}
        &  \makecell[l]{0.951 $\pm$ 0.004}
        &  \makecell[l]{0.587 $\pm$ 0.004}
        &  \makecell[l]{0.693 $\pm$ 0.004}
        &  \makecell[l]{-8.238 $\pm$ 0.101}
        &  \makecell[l]{0.659 $\pm$ 0.014}
        &  \makecell[l]{2.865 $\pm$ 0.038}
        &  \makecell[l]{2.999 $\pm$ 0.163}
        &  \makecell[l]{0.754 $\pm$ 0.005 }
        \\
        \textbf{ }
        &  \textbf{\makecell[l]{\algname (w/o \TOPN )}}
        &  \makecell[l]{0.857 $\pm$ 0.061}
        &  \makecell[l]{0.627 $\pm$ 0.009}
        &  \makecell[l]{0.717 $\pm$ 0.004}
        &  \makecell[l]{-8.583 $\pm$ 0.075}
        &  \makecell[l]{0.727 $\pm$ 0.019}
        &  \makecell[l]{2.566 $\pm$ 0.088}
        &  \makecell[l]{3.388 $\pm$ 0.095}
        &  \makecell[l]{0.717 $\pm$ 0.028 }
        \\
        \textbf{ }
        &  \textbf{\makecell[l]{\algname (w/o APO)}}
        &  \makecell[l]{0.998 $\pm$ 0.001}
        &  \makecell[l]{\underline{0.666} $\pm$ 0.000}
        &  \makecell[l]{\underline{0.740} $\pm$ 0.001}
        &  \makecell[l]{\underline{-9.312} $\pm$ 0.018}
        &  \makecell[l]{\underline{0.734} $\pm$ 0.002}
        &  \makecell[l]{{2.698} $\pm$ 0.006 }
        &  \makecell[l]{\underline{3.676} $\pm$ 0.006 }
        &  \makecell[l]{0.813 $\pm$ 0.002 }
        \\
        \textbf{ }
        &  \textbf{\makecell[l]{\algname}}
        &  \makecell[l]{0.944 $\pm$ 0.094}
        &  \makecell[l]{\textbf{0.675} $\pm$ 0.031}
        &  \makecell[l]{\textbf{0.740} $\pm$ 0.015}
        &  \makecell[l]{\textbf{-9.343} $\pm$ 0.440}
        &  \makecell[l]{\textbf{0.746} $\pm$ 0.028}
        &  \makecell[l]{\underline{2.453} $\pm$ 0.154}
        &  \makecell[l]{\textbf{3.913} $\pm$ 0.358}
        &  \makecell[l]{0.745 $\pm$ 0.032 }
        \\
        \bottomrule
        \textbf{RTCB}
        &  \textbf{\makecell[l]{Original}}
        &  \makecell[l]{-}
        &  \makecell[l]{0.536}
        &  \makecell[l]{{0.698}}
        &  \makecell[l]{-8.572}
        &  \makecell[l]{0.709}
        &  \makecell[l]{3.005}
        &  \makecell[l]{2.299}
        &  \makecell[l]{-}
        \\
        (PDBID:
        &  \textbf{\makecell[l]{MMP \citep{loeffler2024reinvent}}}
        &  \makecell[l]{0.998 $\pm$ 0.001}
        &  \makecell[l]{0.636 $\pm$ 0.000}
        &  \makecell[l]{0.731 $\pm$ 0.001}
        &  \makecell[l]{-8.465 $\pm$ 0.021}
        &  \makecell[l]{0.709 $\pm$ 0.001}
        &  \makecell[l]{2.599 $\pm$ 0.004}
        &  \makecell[l]{3.013 $\pm$ 0.013}
        &  \makecell[l]{0.845 $\pm$ 0.001}
        \\
        \ 4DWQ)
        &  \textbf{\makecell[l]{Similarity ($\geq$ 0.5) \citep{loeffler2024reinvent}}}
        &  \makecell[l]{0.999 $\pm$ 0.001}
        &  \makecell[l]{0.626 $\pm$ 0.000}
        &  \makecell[l]{0.723 $\pm$ 0.001}
        &  \makecell[l]{-8.511 $\pm$ 0.012}
        &  \makecell[l]{0.713 $\pm$ 0.002}
        &  \makecell[l]{2.543 $\pm$ 0.002}
        &  \makecell[l]{3.082 $\pm$ 0.031}
        &  \makecell[l]{0.760 $\pm$ 0.000}
        \\
        \textbf{ }
        &  \textbf{\makecell[l]{Similarity ([0.5, 0.7)]) \citep{loeffler2024reinvent}}}
        &  \makecell[l]{0.999 $\pm$ 0.001}
        &  \makecell[l]{0.622 $\pm$ 0.001}
        &  \makecell[l]{0.718 $\pm$ 0.000}
        &  \makecell[l]{-8.486 $\pm$ 0.021}
        &  \makecell[l]{0.713 $\pm$ 0.003}
        &  \makecell[l]{2.542 $\pm$ 0.005}
        &  \makecell[l]{3.101 $\pm$ 0.005}
        &  \makecell[l]{0.740 $\pm$ 0.001}
        \\
        \textbf{ }
        &  \textbf{\makecell[l]{Similarity ($\geq$ 0.7) \citep{loeffler2024reinvent}}}
        &  \makecell[l]{0.999 $\pm$ 0.001}
        &  \makecell[l]{0.639 $\pm$ 0.000}
        &  \makecell[l]{0.734 $\pm$ 0.001}
        &  \makecell[l]{-8.496 $\pm$ 0.009}
        &  \makecell[l]{ 0.718 $\pm$ 0.001}
        &  \makecell[l]{2.628 $\pm$ 0.001}
        &  \makecell[l]{2.868 $\pm$ 0.003}
        &  \makecell[l]{0.875 $\pm$ 0.002}
        \\
        \textbf{ }
        &  \textbf{\makecell[l]{Scaffold \citep{loeffler2024reinvent}}}
        &  \makecell[l]{0.998 $\pm$ 0.001}
        &  \makecell[l]{0.609 $\pm$ 0.001}
        &  \makecell[l]{0.718 $\pm$ 0.000}
        &  \makecell[l]{-8.508 $\pm$ 0.026}
        &  \makecell[l]{0.711 $\pm$ 0.000}
        &  \makecell[l]{2.627 $\pm$ 0.002}
        &  \makecell[l]{ 2.803 $\pm$ 0.010}
        &  \makecell[l]{0.735 $\pm$ 0.002}
        \\
        \textbf{ }
        &  \textbf{\makecell[l]{Scaffold Generic \citep{loeffler2024reinvent}}}
        &  \makecell[l]{0.998 $\pm$ 0.001}
        &  \makecell[l]{0.625 $\pm$ 0.001}
        &  \makecell[l]{0.722 $\pm$ 0.000}
        &  \makecell[l]{-8.544 $\pm$ 0.009}
        &  \makecell[l]{0.722 $\pm$ 0.002}
        &  \makecell[l]{2.551 $\pm$ 0.010}
        &  \makecell[l]{2.898 $\pm$ 0.005}
        &  \makecell[l]{0.768 $\pm$ 0.004}
        \\
        \textbf{ }
        &  \textbf{\makecell[l]{DrugImprover \citep{liu2023drugimprover}}}
        &  \makecell[l]{0.920 $\pm$ 0.008}
        &  \makecell[l]{0.478 $\pm$ 0.001} 
        &  \makecell[l]{0.618 $\pm$ 0.002} 
        &  \makecell[l]{-8.701 $\pm$ 0.037}
        &  \makecell[l]{0.486 $\pm$ 0.002}
        &  \makecell[l]{\textbf{1.181} $\pm$ 0.010}
        &  \makecell[l]{2.026 $\pm$ 0.013}
        &  \makecell[l]{0.427 $\pm$ 0.001}
        \\
        \textbf{ }
        &  \textbf{\makecell[l]{Molsearch \citep{sun2022molsearch}}}
        &  \makecell[l]{\textbf{1.000} $\pm$ 0.001}
        &  \makecell[l]{0.625 $\pm$ 0.001} 
        &  \makecell[l]{0.742 $\pm$ 0.001} 
        &  \makecell[l]{-8.747 $\pm$ 0.009}
        &  \makecell[l]{0.719 $\pm$ 0.001}
        &  \makecell[l]{3.012 $\pm$ 0.004}
        &  \makecell[l]{2.273 $\pm$ 0.005}
        &  \makecell[l]{\textbf{0.950} $\pm$ 0.001}
        \\
        \textbf{ }
        &  \textbf{\makecell[l]{MIMOSA \citep{fu2021mimosa}}}
        &  \makecell[l]{0.989 $\pm$ 0.001}
        &  \makecell[l]{0.631 $\pm$ 0.001} 
        &  \makecell[l]{0.749 $\pm$ 0.001} 
        &  \makecell[l]{-8.972 $\pm$ 0.011}
        &  \makecell[l]{0.706 $\pm$ 0.003}
        &  \makecell[l]{3.080 $\pm$ 0.007}
        &  \makecell[l]{2.561 $\pm$ 0.008}
        &  \makecell[l]{\underline{0.945} $\pm$ 0.001}
        \\
        \textbf{ }
        &  \textbf{\makecell[l]{\fix{DrugEx v3} \citep{liu2023drugex}}}
        &  \makecell[l]{\textbf{1.000} $\pm$ 0.001}
        &  \makecell[l]{0.592 $\pm$ 0.001} 
        &  \makecell[l]{0.668 $\pm$ 0.001} 
        &  \makecell[l]{-8.762 $\pm$ 0.010}
        &  \makecell[l]{0.583 $\pm$ 0.002}
        &  \makecell[l]{\underline{2.488} $\pm$ 0.005}
        &  \makecell[l]{5.827 $\pm$ 0.010}
        &  \makecell[l]{0.393 $\pm$ 0.001}
        \\
        \textbf{ }
        &  \textbf{\makecell[l]{\algname (w/o APO \& \TOPN )}}
        &  \makecell[l]{0.956 $\pm$ 0.004}
        &  \makecell[l]{0.582 $\pm$ 0.007}
        &  \makecell[l]{0.700 $\pm$ 0.008}
        &  \makecell[l]{-8.214 $\pm$ 0.125}
        &  \makecell[l]{0.686 $\pm$ 0.017}
        &  \makecell[l]{2.788 $\pm$ 0.056}
        &  \makecell[l]{2.781 $\pm$ 0.214}
        &  \makecell[l]{0.707 $\pm$ 0.005 }
        \\
        \textbf{ }
        &  \textbf{\makecell[l]{\algname (w/o \TOPN)}}
        &  \makecell[l]{0.811 $\pm$ 0.074}
        &  \makecell[l]{0.639 $\pm$ 0.004}
        &  \makecell[l]{0.723 $\pm$ 0.005}
        &  \makecell[l]{-8.808 $\pm$ 0.071}
        &  \makecell[l]{0.741 $\pm$ 0.013}
        &  \makecell[l]{{2.521} $\pm$ 0.081}
        &  \makecell[l]{3.279 $\pm$ 0.067}
        &  \makecell[l]{0.730 $\pm$ 0.030 }
        \\
        \textbf{ }
        &  \textbf{\makecell[l]{\algname (w/o APO)}}
        &  \makecell[l]{0.997 $\pm$ 0.001}
        &  \makecell[l]{\underline{0.673} $\pm$ 0.001}
        &  \makecell[l]{\underline{0.755} $\pm$ 0.001}
        &  \makecell[l]{\underline{-9.659} $\pm$ 0.023}
        &  \makecell[l]{\underline{0.764} $\pm$ 0.001}
        &  \makecell[l]{{2.606} $\pm$ 0.007}
        &  \makecell[l]{\underline{3.481} $\pm$ 0.027}
        &  \makecell[l]{0.773 $\pm$ 0.003 }
        \\
        \textbf{ }
        &  \textbf{\makecell[l]{\algname}}
        &  \makecell[l]{0.826 $\pm$ 0.100}
        &  \makecell[l]{\textbf{0.682} $\pm$ 0.004}
        &  \makecell[l]{\textbf{0.756} $\pm$ 0.003}
        &  \makecell[l]{\textbf{-9.757} $\pm$ 0.057}
        &  \makecell[l]{\textbf{0.765} $\pm$ 0.013}
        &  \makecell[l]{{2.437} $\pm$ 0.059}
        &  \makecell[l]{\textbf{3.582} $\pm$ 0.043}
        &  \makecell[l]{0.747 $\pm$ 0.026 }
        \\
        \bottomrule
        \\
        \end{tabular}}}
        \caption{
        {\textbf{Main results.} A comparison of eight baselines including Original, six baselines from REINVENT \{MMP, Similarity ($\geq 0.5$), Similarity $\in [0.5,0.7)$, Similarity $\geq 0.7$, Scaffold, Scaffold Generic\}, DrugImprover, Molsearch, MIMOSA and different versions of \algname on various objectives  
        based on 3CLPro and RTCB datasets. {The top two results are highlighted as \textbf{1st} and \underline{2nd}. $^\star$ represents the top-priority target objective.} Results are reported for five experimental runs. 
        }
        }
        \label{exp:main_result}
\end{table*}

\subsection{Main results.}
\XL{add molecule}

Table \ref{exp:main_result} shows that \algname surpasses DrugImprover and six different versions of REINVENT4 in performance measures for both virus-related and cancer-related proteins. Moreover, \algname exceeds the performance of all baseline methods and also demonstrates a decent level of Tanimoto similarity to the original drug, indicating that it preserves the advantageous features of the original drugs while improving desired properties.

Several key factors contribute to this superior performance. Although DrugImprover established a strong foundation for the drug optimization field, including a workflow and a reinforcement learning algorithm to align the generative model with multiple pharmaceutical objectives, \algname outshines DrugImprover in all benchmarks. This is because \algname employs a GPT-2-like Transformer as the basis of its generative model, whereas DrugImprover relies solely on LSTM. Consequently, the GPT-2 Transformer grants \algname enhanced scalability, capacity, and contextual understanding compared to DrugImprover.

In contrast to the current state-of-the-art approach, REINVENT4, which pre-trains a Transformer with constraints on Tanimoto similarity, their method falls short in achieving drug optimization as it overlooks the optimization of multiple pharmaceutical properties. Therefore, Table \ref{exp:main_result} reveals that although REINVENT4 achieved high similarity, the generated molecules often failed to surpass the original ones.
\algname, on the other hand, employs the APO reinforcement learning algorithm to fine-tune the pre-trained generative model and utilizes the $\TOPN$ decoding optimization strategy. These approaches ensure improvements aligned with multiple pharmaceutical objectives and enable \algname to successfully enhance the original drug across various pharmaceutical properties while maintaining a high Tanimoto similarity.

\subsection{Ablation studies.}

\begin{table*}[ht]
\setlength{\tabcolsep}{4pt}
   \centering
\scalebox{0.44}{
\begin{tabular}{|c|c|c|c|c|c|c|c|c|c|}
\hline
\textbf{Model} & \textbf{Diversity}~$\uparrow$ & \textbf{Validity}~$\uparrow$ & \textbf{Avg Norm Reward}~$\uparrow$$^\star$ & \textbf{ \makecell[c]{Avg Top 10\% \\Norm Reward}}~$\uparrow$ & \textbf{Docking}~$\downarrow$  & \textbf{Druglikeliness}~$\uparrow$ & \textbf{Synthesizability}~$\downarrow$ & \textbf{Solubility}~$\uparrow$ & \textbf{Similarity}~$\uparrow$ \\
\hline
\multicolumn{10}{|c|}{\textbf{COVID}} \\
\hline
\algname (w/o \TOPN) \TOPK & 0.988 & \textbf{0.947} & 0.645 & 0.727 & -8.726 & \textbf{0.760} & 2.418 & 3.499 & 0.697 \\
\algname (w/o \TOPN) \TOPP & 0.988 & 0.938 & 0.642 & 0.722 & -8.653 & 0.756 & 2.420 & 3.505 & 0.696 \\
\algname (w/o \TOPN) \TOPPK & \textbf{0.989} & 0.941 & 0.643 & 0.724 & -8.667 & 0.759 & \textbf{2.407} & 3.506 & 0.692 \\
\algname (\TOPN) & 0.965 & 0.944 & \textbf{0.675} & \textbf{0.740} & \textbf{-9.343} & {0.746} & {2.453} & \textbf{3.913} & \textbf{0.745} \\
\hline
\multicolumn{10}{|c|}{\textbf{CANCER}} \\
\hline
\algname (w/o \TOPN) \TOPK & 0.912 & 0.709 & 0.648 & 0.728 & -8.944 & 0.756 & 2.456 & 3.258 & 0.730 \\
\algname (w/o \TOPN) \TOPP & \textbf{0.931} & 0.704 & 0.645 & 0.729 & -8.907 & 0.756 & 2.466 & 3.226 & 0.723 \\
\algname (w/o \TOPN) \TOPPK & 0.926 & 0.719 & 0.645 & 0.727 & -8.888 & 0.757 & 2.466 & 3.219 & 0.725 \\
\algname (\TOPN) & 0.916 & \textbf{0.826} & \textbf{0.682} & \textbf{0.756} & \textbf{-9.757} & \textbf{0.765} & \textbf{2.437} & \textbf{3.582} & \textbf{0.747} \\
\hline
\end{tabular}
}
\caption{Ablation study of \TOPN, \TOPP, \TOPK and \TOPPK sampling strategies. The top result is highlighted as \textbf{1st}.  $^\star$ represents the top-priority target objective. \fix{Evaluations are based on five random seeds.} \TOPN outperforms others in most of the metrics.} 
\label{tab:stage-ablation-topn}
\end{table*}

We next present ablation studies that underscore the necessity and effectiveness of each component of \algname. These components complement each other, substantially enhancing overall performance.

\begin{wrapfigure}[9]{r}{.26\textwidth}
    \centering
\setlength{\tabcolsep}{4pt}
   \centering
\scalebox{0.51}{
\begin{tabular}{|c|c|}
\hline
\textbf{1024 SMILES} & \textbf{Details} \\
\hline
1024 & $>100$ length, $>50$ scaffold \\
\hline
\multicolumn{2}{|c|}{\textbf{Validity}} \\
\hline
One phase & 0.57 \fix{$\pm$ 0.010} \\
Two phases & 0.68 \fix{$\pm$ 0.006} \\
\hline
\end{tabular}}
        \caption{Ablation study of one-phase vs two-phases. \fix{Evaluations are based on five random seeds.}} 
        \label{tab:stage-ablation}
\end{wrapfigure}
\paragraph{Effectiveness of (two-phase) incremental training.}
In two-phase incremental pretraining, 
the intuition behind the first phase lies in training on critical keywords as knowledge pieces, reinforcing the memory of these key terms, particularly in longer sequences.
We conducted an ablation study comparing our novel two-phase incremental training with \fix{conventional} one-phase training in \tabref{tab:stage-ablation}.
\fix{To ensure a fair comparison in terms of total training epochs, we trained for 10 epochs in the conventional single-phase setting, and for five epochs per phase in our two-phase setting.}
The results showed that the two-phase approach improves validity compared to one-phase training, demonstrating the effectiveness of the incremental training method.

\paragraph{Effectiveness of APO Finetuning.} \algname adopts APO finetuning as the second step, following the completion of pretraining the GPT-based generator. \tabref{exp:main_result} demonstrates the effectiveness of APO through two comparisons: \algname (w/o APO, \TOPN) vs. \algname (w/o \TOPN), which shows that after applying APO finetuning, performance improved on most properties. Additionally, \algname vs. \algname (w/o APO) validates the importance of the APO component. By applying APO on top of pretraining and \TOPN decoding, performance improved. Both cases demonstrate the effectiveness of APO finetuning. \fix{ Note that APO may compromise certain reward metrics, such as similarity or validity, if this trade-off leads to improved performance in the target weighted objective.
}

\paragraph{Effectiveness of $\TOPN$ decoding strategy.} \algname adopts the $\TOPN$ decoding strategy as the final step followed by APO finetuning. \tabref{exp:main_result} demonstrates the effectiveness of $\TOPN$ through two comparisons: \algname (w/o APO, $\TOPN$) vs. \algname (w/o APO), showing that after applying the $\TOPN$ decoding strategy on top of pretrained GPT, performance improved across most properties.
Moreover, \algname vs. \algname (w/o $\TOPN$) illustrates that after applying APO on top of pretraining and RL, performance still improves on multiple attributes, surpassing all baselines.
Furthermore, by comparing \algname (w/o APO) and \algname (w/o $\TOPN$), we observe that applying $\TOPN$ decoding alone enhances performance more than applying APO alone.

\begin{table*}[t]
\setlength{\tabcolsep}{4pt}
   \centering
    {\small
    \scalebox{0.56}{
    \begin{tabular}{ l l l l l l}
        \toprule
        \textbf{\makecell[l]{ }}
        &  \makecell[l]{\textbf{Original~~~~~~~~~~~~~Scaffold}}
        &  \makecell[l]{\textbf{REINVENT}}
        &  \makecell[l]{\textbf{\algname 1}}
        &  \makecell[l]{\textbf{\algname 2}}
        &  \makecell[l]{\textbf{\algname 3}}
        \\
        \textbf{Molecule}
        & {\makecell[l]{\includegraphics[width=0.35\textwidth]{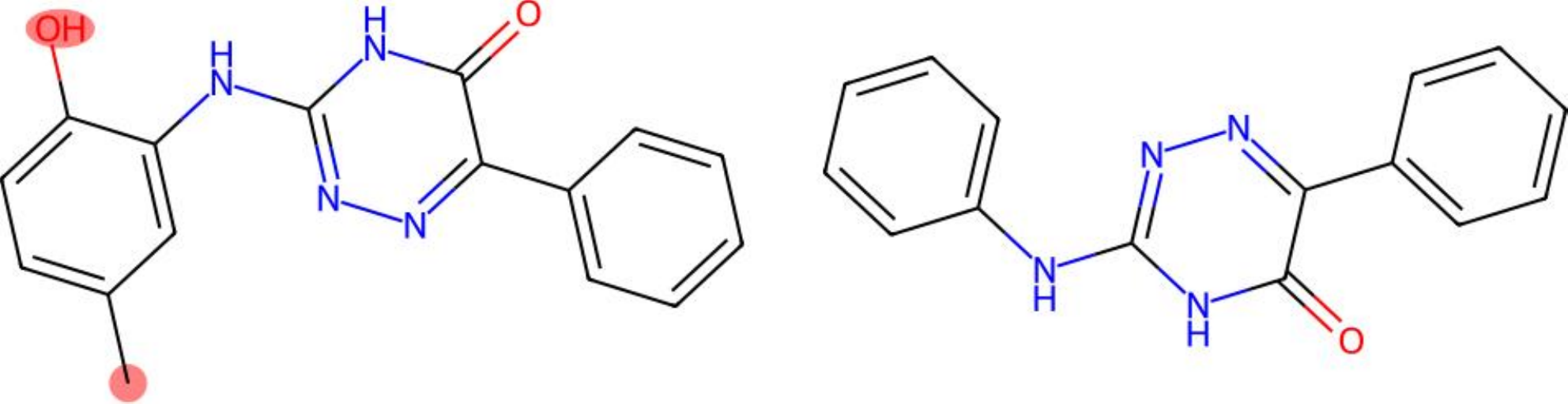}}}
        & {\makecell[l]{\includegraphics[width=0.25\textwidth]{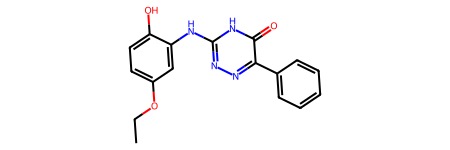}}}
        & {\makecell[l]{\includegraphics[width=0.25\textwidth]{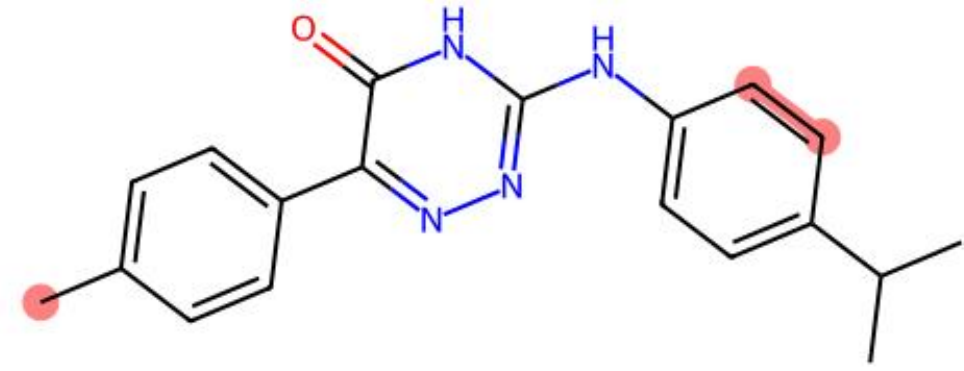}}}
        & {\makecell[l]{\includegraphics[width=0.25\textwidth]{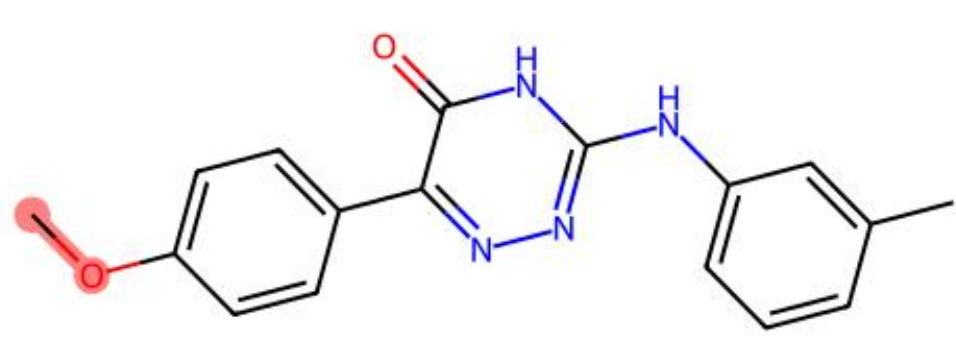}}}
        & {\makecell[l]{\includegraphics[width=0.25\textwidth]{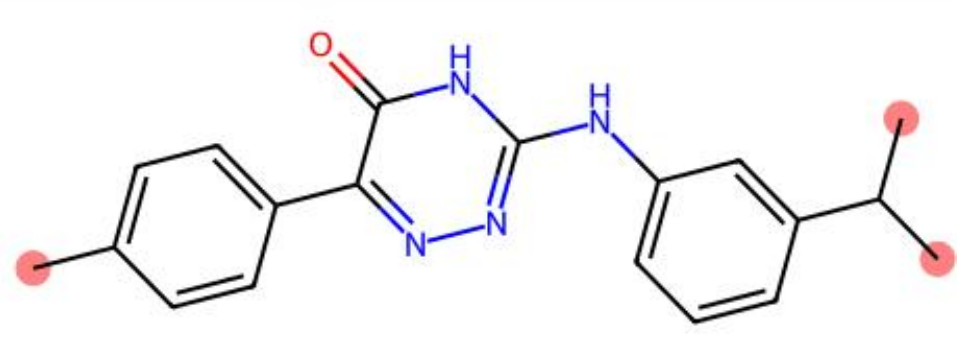}}}
        \\
        \midrule
        \textbf{\makecell[l]{SMILE String}}
        &  \makecell[l]{Cc1ccc(O)c(Nc2nnc\\(-c3ccccc3)c(=O)[nH]2)c1}
        &  \makecell[l]{CCOc1ccc(O)c(Nc2nnc\\(-c3ccccc3)c(=O)[nH]2)c1}
        &  \makecell[l]{Cc1ccc(-c2nnc(Nc3ccc\\(C(C)C)cc3)[nH]c2=O)cc1} 
        &  \makecell[l]{COc1ccc(-c2nnc(Nc3cccc\\(C)c3)[nH]c2=O)cc1} 
        &  \makecell[l]{Cc1ccc(-c2nnc(Nc3cccc\\(C(C)C)c3)[nH]c2=O)cc1}
        \\
        \midrule
        \textbf{\makecell[l]{Scaffold}}
        &  \makecell[l]{O=c1[nH]c(Nc2cc\\ccc2)nnc1-c1ccccc1}
        &  \makecell[l]{c1ccc(Nc2nnc(-c3\\ccccc3)[nH]2)cc1}
        &  \makecell[l]{same as original} 
        &  \makecell[l]{same as original} 
        &  \makecell[l]{same as original}
        \\
        \textbf{\makecell[l]{Similarity}}
        &  \makecell[l]{-}
        &  \makecell[l]{0.878} 
        &  \makecell[l]{0.826} 
        &  \makecell[l]{0.911} 
        &  \makecell[l]{0.866}
        \\
        \midrule
        \textbf{\makecell[l]{Docking~$(\downarrow)$}}
        &  \makecell[l]{-10.031}
        &  \makecell[l]{-9.258}
        &  \makecell[l]{-11.478 $\checkmark$} 
        &  \makecell[l]{-11.474 $\checkmark$} 
        &  \makecell[l]{-11.087 $\checkmark$}
        \\
        \textbf{\makecell[l]{Druglikeness~$(\uparrow)$}}
        &  \makecell[l]{0.646}
        &  \makecell[l]{0.624}
        &  \makecell[l]{0.762 $\checkmark$}
        &  \makecell[l]{0.774 $\checkmark$}
        &  \makecell[l]{0.762 $\checkmark$}
        \\
         \textbf{\makecell[l]{Synthesizability~$(\downarrow)$}}
        &  \makecell[l]{2.390}
        &  \makecell[l]{2.417} 

        &  \makecell[l]{2.298 $\checkmark$} 
        &  \makecell[l]{2.257 $\checkmark$} 
        &  \makecell[l]{2.356 $\checkmark$}
        \\
         \textbf{\makecell[l]{Solubility~$(\uparrow)$}}
        &  \makecell[l]{2.590} 
        &  \makecell[l]{2.680 $\checkmark$} 
        &  \makecell[l]{4.007 $\checkmark$} 
        &  \makecell[l]{2.893 $\checkmark$} 
        &  \makecell[l]{4.007 $\checkmark$} 
        \\
        \textbf{\makecell[l]{Avg Norm Reward~$(\uparrow)$$^\star$}}
        &  \makecell[l]{0.618}
        &  \makecell[l]{0.589}        
        &  \makecell[l]{0.759 $\checkmark$} 
        &  \makecell[l]{0.753 $\checkmark$} 
        &  \makecell[l]{0.754 $\checkmark$}
        \\
        \bottomrule
    \end{tabular}}}
        \caption{One optimization example from cancer benchmark. Every generated molecules retains the scaffold, with all desired properties improved compared to the original. $^\star$ represents the top-priority target objective. $\checkmark$ indicates improved property. 
        }
        \label{viz:example1}
\end{table*}

\fix{\paragraph{Top-N selects a higher‑reward alternative missed by Top-K.} These two SMILES strings are nearly identical, but TOP-N selects a slightly different generation path—preferring ‘c’ and ‘c1C’ over ‘c(C)c1’—which results in a higher average reward. This demonstrates how TOP-N can identify token-level alternatives that improve the optimization objective while preserving scaffold structure.}

\begin{table}[ht]
\centering
\scalebox{0.85}{
\begin{tabular}{|l|l|c|}
\hline
\textbf{Method} & \textbf{Generated SMILES} & \makecell{\textbf{Avg.} \\ \textbf{Normalized Reward}} \\
\hline
Top-K & Cc1ccc(NC(=O)COC(=O)C2C[C@H]3CCCC@HC3=O)c(C)c1 & 0.579 \\
Top-N & Cc1cccc(NC(=O)COC(=O)C2C[C@H]3CCCC@HC3=O)c1C & 0.614 \\
\hline
\end{tabular}
}
\caption{\fix{Comparison of generated SMILES and average normalized rewards across methods.}}
\end{table}

\fix{
\paragraph{Ablation study of \TOPN vs \TOPP, \TOPK and \TOPPK strategies.}
\fix{Following the setup described in \secref{exp_config}, we perform an ablation study on sampling strategies.}
When removing \TOPN component, we employ multinomial generation, where multinomial sampling randomly selects the next token from the entire vocabulary based on the model's probability distribution. In the ablation study detailed in \tabref{tab:stage-ablation-topn}, we examined \TOPN over \TOPP, \TOPK, and \TOPPK  with K=20 and P=0.95. The results indicate that Top-N surpasses the other strategies in most metrics.}

\paragraph{Drug optimization illustration.} 
Finally, we provide three examples illustrating the effectiveness of \algname in improving upon the original molecule on the cancer benchmark, as shown in \tabref{viz:example1} (Refer to \appref{app:example_2} for the COVID benchmark). The results in \tabref{viz:example1} demonstrate that the drugs generated by \algname outperform the original drugs across all desired properties. Additionally, the comparison figure in \tabref{viz:example1} illustrates that the improved molecules preserve the original drug to a significant extent, with only minor changes highlighted in {\color{myred}red}. The results indicate that \algname effectively optimizes desired properties while preserving the beneficial properties of the original drug.

\section{Conclusion}\label{sec:con}

We have introduced \algname, a novel framework for drug optimization. This framework incorporates a unique scaffold-based GPT design, a three-stage optimization process, a two-phase incremental pre-training approach, and a novel \TOPN decoding strategy that facilitates controlled reward-guided generation using GPT. To showcase the superior performance of \algname, we conduct evaluations on real-world viral and cancer-related datasets to compare it against eight competing baselines.
Our results demonstrate that \algname surpasses all baselines across the majority of performance metrics, underscoring its efficacy. 
Our work highlights \algname's effectiveness in drug optimization, as evidenced by enhancements in various pharmaceutical properties. 
\fix{Currently, \algname is limited to handling molecules in SMILES format. We are working to expand \algname's capabilities to accommodate a broader range of drug representation formats as a future direction.}

\subsubsection*{Acknowledgements}
This work is supported in part by the RadBio-AI project (DE-AC02-06CH11357), U.S. Department of Energy Office of Science, Office of Biological and Environment Research, the Improve project under contract (75N91019F00134, 75N91019D00024, 89233218CNA000001, DE-AC02-06-CH11357, DE-AC52-07NA27344, DE-AC05-00OR22725), 
the Exascale Computing Project (17-SC-20-SC), a collaborative effort of the U.S. Department of Energy Office of Science and the National Nuclear Security Administration.

\bibliography{reference}

\clearpage
\onecolumn

\appendix

 \clearpage

\section{Appendix}

\subsection{Pre-training and finetuning dataset}\label{app:pretrain_data}
For pretrainig, We used the ZINC dataset, filtering for Standard, In-Stock, and Drug-Like molecules, resulting in approximately 11 million molecules. 
\fix{
For preprocessing, we perform a few straightforward steps: \\
1. We first canonicalize smiles
{Chem.MolToSmiles(Chem.MolFromSmiles(mol),True)}. \\
2. We filter out molecules whose scaffold SMILES is an empty string. These preprocessing steps are also included in the huggingface data repo.
}

\fix{For finetuning, we utilize 1 million compounds from the ZINC15 dataset, docked to the 3CLPro protein (PDB ID: 7BQY) linked to SARS-CoV-2 and the RTCB protein (PDB ID: 4DWQ) associated with human cancer, as sourced from the latest Cancer and COVID dataset by \citet{liu2023drugimprover}, across all baselines.}

{
\subsection{Generation with finetuned model}\label{app:generation}
The top five epochs with the highest historical average normalized reward (as detailed in Section \ref{exp_config}) are selected. From these five epochs, the epoch with the highest product of validity and average normalized reward is chosen as the final model for generation.

With this epoch and corresponding weights, we apply the proposed decoding method (as described in section \ref{decoding_strategy}) for generation. 
}

\subsection{{BPE Tokenization}}\label{app:BPE}
The Byte Pair Encoding (BPE) algorithm involves the following steps:
\begin{enumerate}[noitemsep, topsep=0pt, partopsep=0pt]
    \item \textbf{Initialize the Vocabulary:} Start with a base vocabulary consisting of all individual characters in the text corpus.
    \item \textbf{Count Frequencies:} Count the frequency of all character pairs in the text.
    \item \textbf{Merge Most Frequent Pair:} Identify the most frequent pair of characters and merge them into a single token. Add this new token to the vocabulary.
    \item \textbf{Update Text:} Replace all occurrences of the most frequent pair with the new token in the text.
    \item \textbf{Repeat:} Repeat the process of counting frequencies, merging pairs, and updating the text until the desired vocabulary size is reached or no more merges are possible.
\end{enumerate}
BPE constructs a robust vocabulary by iteratively merging the most frequent token pairs, effectively capturing common subword units for more efficient and flexible text representation. The resulting vocabulary comprises 3,152 tokens and includes special tokens as well. For instance, the sequence $\textsc{<L>}$ is tokenized into three separate tokens: $<$, L, and $>$. The tokenizer was trained on 10 million molecules from the ZINC dataset, ensuring comprehensive coverage of chemical elements.

\subsection{Surrogate model}\label{app:surrogate_model}

{The surrogate model~\citep{vasan23} is a simplified variant of a BERT-like transformer, extensively utilized in natural language processing. In this model, tokenized SMILES strings are inputted and then embedded with positional information. The resulting outputs are subsequently fed into a series of five transformer blocks, each comprising a multi-head attention layer (21 heads), a dropout layer, layer normalization with residual connection, and a feedforward network. This feedforward network consists of two dense layers followed by dropout and layer normalization with residual connection. Following the stack of transformer blocks, a final feedforward network is employed to generate the predicted docking score.}
\fix{The validation $r^2$ values are 0.842 for 3CLPro and 0.73 for the RTCB dataset.}

\subsection{Drug Optimization illustration on COVID benchmark}\label{app:example_2}
{This is another example illustrating the effectiveness of \algname in enhancing the original molecule on the COVID benchmark. The results in \tabref{viz:example2} show that the drugs generated by \algname outperform the original drugs across all desired properties. Even though the original scaffold is altered and not present in the generated molecules, the similarity still demonstrates a decent level.}

\begin{table*}[t!]
\setlength{\tabcolsep}{4pt}
   \centering
    {\small
    \scalebox{0.75}{
    \begin{tabular}{ l l l l }
        \toprule
        \textbf{\makecell[l]{ }}
        &  \makecell[l]{\textbf{Original~~~~~~~~~~~~~~~~~~~~~~~~Scaffold}}
        &  \makecell[l]{\textbf{improved 1}}
        &  \makecell[l]{\textbf{improved 2}}
        \\
        \midrule
        \textbf{Molecule}
        & {\makecell[l]{\includegraphics[width=0.3\textwidth]{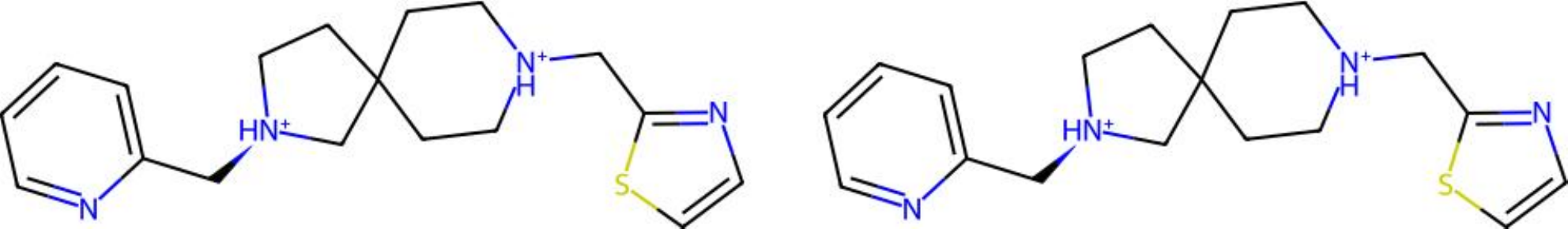}}}
        & {\makecell[l]{\includegraphics[width=0.3\textwidth]{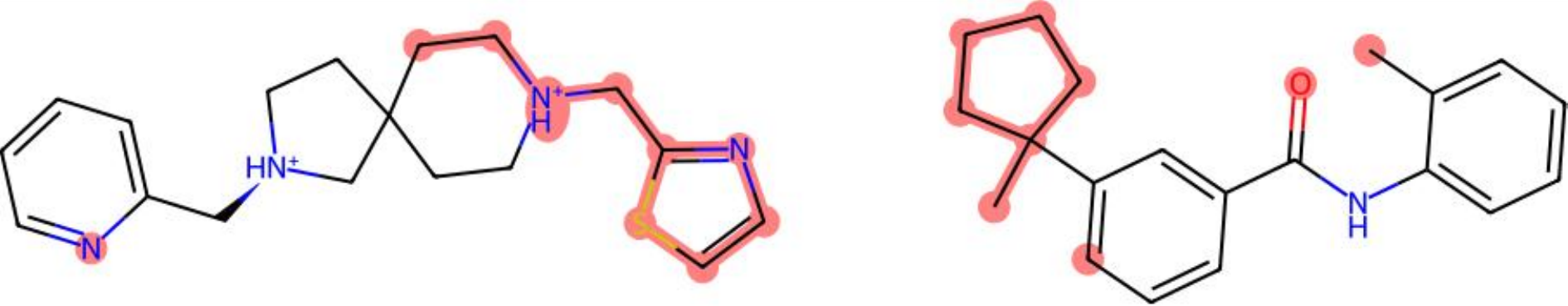}}}
        & {\makecell[l]{\includegraphics[width=0.3\textwidth]{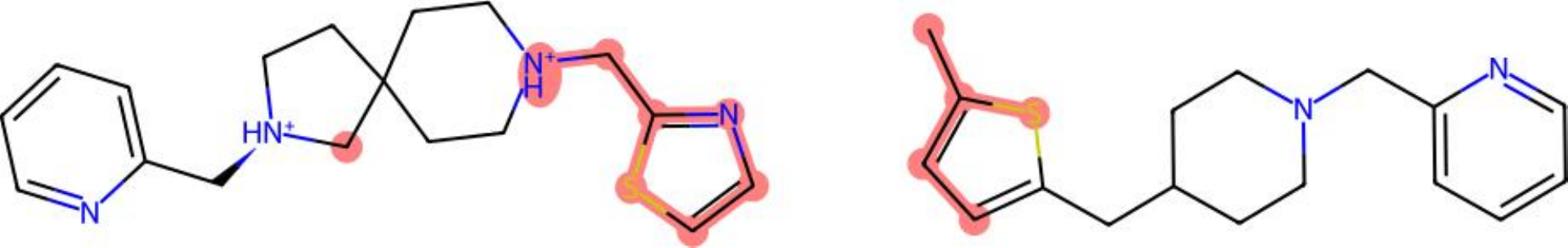}}}
        \\
        \midrule
        \textbf{\makecell[l]{SMILE String}}
        &  \makecell[l]{c1ccc(C[N@H+]2CCC3(CC[NH+]\\(Cc4nccs4)CC3)C2)nc1}
        &  \makecell[l]{CC1(c2cccc(C(=O)\\Nc3ccccc3C)c2)CCCC1} 
        &  \makecell[l]{Cc1ccc(CC2CCN\\(Cc3ccccn3)CC2)s1} 
        \\
        \midrule
        \textbf{\makecell[l]{Scaffold}}
        &  \makecell[l]{c1ccc(C[N@H+]2CCC3(CC[NH+]\\(Cc4nccs4)CC3)C2)nc1}
        &  \makecell[l]{-} 
        &  \makecell[l]{-} 
        \\
        \midrule
        \textbf{\makecell[l]{Docking~$(\downarrow)$}}
        &  \makecell[l]{-9.748}
        &  \makecell[l]{-10.184~$\checkmark$} 
        &  \makecell[l]{-10.187~$\checkmark$} 
        \\
        \textbf{\makecell[l]{Druglikeness~$(\uparrow)$}}
        &  \makecell[l]{0.839}
        &  \makecell[l]{0.840~$\checkmark$}
        &  \makecell[l]{0.847~$\checkmark$}
        \\
         \textbf{\makecell[l]{Synthesizability~$(\downarrow)$}}
        &  \makecell[l]{5.631}
        &  \makecell[l]{1.983~$\checkmark$} 
        &  \makecell[l]{2.199~$\checkmark$} 
        \\
         \textbf{\makecell[l]{Solubility~$(\uparrow)$}}
        &  \makecell[l]{0.192} 
        &  \makecell[l]{5.079~$\checkmark$} 
        &  \makecell[l]{3.906~$\checkmark$} 
        \\
        \textbf{\makecell[l]{Similarity}}
        &  \makecell[l]{-}
        &  \makecell[l]{0.335} 
        &  \makecell[l]{0.563} 
        \\
        \textbf{\makecell[l]{Avg Norm Reward~$(\uparrow)$}}
        &  \makecell[l]{0.398}
        &  \makecell[l]{0.688~$\checkmark$} 
        &  \makecell[l]{0.694~$\checkmark$} 
        \\
        \bottomrule
    \end{tabular}}}
        \caption{One molecule example from 3CLPro dataset, where scaffold and original are same. In this case the model tries to modify the scaffold, and the generated molecules does not contain scaffold. $\checkmark$ indicates improved property. 
        }
        \label{viz:example2}
\end{table*}

\subsection{{Computing infrastructure {and wall-time comparison}}}\label{app:computing_infrastructure}
We trained our docking surrogate models using 4 nodes of the supercomputer 
where each node contains CPUs (64 cores) and 4 A100 GPU nodes~\citep{Polaris}.\XL{need update}
The training time for each model was approximately 3 hours.

We conducted other experiments on a cluster that includes CPU nodes (approximately 280 cores) and GPU nodes (approximately 110 Nvidia GPUs, ranging from Titan X to A6000, set up mostly in 4- and 8-GPU configurations). 

{The pretraining process utilizes 8 GPUs, while APO and generation employs a single GPU. Both processes use either V100 or A100 GPUs. Based on the computing infrastructure, we obtained the wall-time comparison in \tabref{table:wall-time} as follows.}

 \begin{table*}[ht!]
 {
    \centering
    {\scriptsize
    \scalebox{1}{
    \begin{tabular}{l c c  }
        \toprule
        \textbf{Methods}
        & {\makecell[c]{Total Run Time}}
        \\
        \midrule
        \textbf{\makecell[l]{Pretraining}}
        &  \makecell[r]{24h}
        \\
        \textbf{\makecell[l]{APO}}
        &  \makecell[r]{27h}
        \\
        \textbf{\makecell[l]{\TOPN (One Generation)}}
        &  \makecell[r]{17-20s}
        \\
        \bottomrule
    \end{tabular}}}
    \caption{{Wall-time comparison between different methods.} }
        \label{table:wall-time}   
        }
\end{table*}

\subsection{{Hyperparameters and architectures}}\label{app:hyperparameters}
Table \ref{app:tab:hyperparams_pretrain} and  \ref{app:tab:hyperparams} provides a list of hyperparameter settings we used for our experiments.\XL{need update}

For APO finetuning and experimentation, \fix{we have two different set of} 1280 molecules that were selected from each of the RTCB and 3CLPro datasets, with docking scores ranging from -14 to -6. This range is based on \citep{liu2024erp}.

{Moreover, when computing the average normalized reward for the original molecule, in the absence of similarity considerations, we use weights of $0.25$ for docking, drug-likeness, synthesizability, and solubility, respectively.
}

{
Moreover, when the generated SMILES is invalid, indicating that the reward $R_c$ cannot be calculated, we have two options: either directly subtract the reward of the original SMILES (i.e., $-R_c(X)$), or consider the advantage preference as zero instead.
}
\begin{table*}[h!]
    {
    \centering
    {\scriptsize
    \scalebox{1}{
    \begin{tabular}{c c }
        \toprule
        \textbf{Parameter} &  \textbf{Value} 
        \\
        \midrule
        {\makecell[l]{Pretraining}}
        \\
        \midrule
        {\makecell[l]{\quad Learning rate}} &  \makecell[c]{$5 \times e^{-5}$}
        \\
        \midrule
        {\makecell[l]{\quad Batch size}} &  \makecell[c]{$24$}
        \\
        \midrule
        {\makecell[l]{\quad Optimizer}} &  \makecell[c]{Adam}
        \\
        \midrule
        {\makecell[l]{\quad \# of Epochs for Training First Phase}} &  \makecell[c]{$10$} \\
        \midrule
        {\makecell[l]{\quad \# of Epochs for Training Second Phase}} &  \makecell[c]{$10$} \\
        \midrule
        {\makecell[l]{\quad Model \# of Params}} &  \makecell[c]{$124M$} 
        \\
        \midrule
        {\makecell[l]{Generation}}
        \\
        \midrule
        {\makecell[l]{\quad N (Top-N)}} &  \makecell[c]{$1$}
        \\
        \midrule
        {\makecell[l]{\quad K (Number of possible next token)}} &  \makecell[c]{$16$}
        \\
        \midrule
        {\makecell[l]{\quad TopK}} &  \makecell[c]{$[10,15,20]$}
        \\
        \midrule
        {\makecell[l]{\quad TopP}} &  \makecell[c]{[$0.85$, $0.9$, $0.95$]}
        \\

        \bottomrule
    \end{tabular}}}
        \caption{{{Hyperparameters for pretraining and generation}}. }
        \label{app:tab:hyperparams_pretrain}
        }
\end{table*}

\begin{table*}[h!]
    {
    \centering
    {\scriptsize
    \scalebox{0.7}{
    \begin{tabular}{c c }
        \toprule
        \textbf{Parameter} &  \textbf{Value} 
        \\
        \midrule
        {\makecell[l]{Shared}}
        \\
        \midrule
        {\makecell[l]{\quad \# of Molecules Optimized}} &  \makecell[c]{$1280$}
        \\
        \midrule
        {\makecell[l]{\quad Learning rate}} &  \makecell[c]{$1 \times 10^{-4}$}
        \\
        \midrule
        {\makecell[l]{\quad Optimizer}} &  \makecell[c]{Adam}
        \\
        \midrule
        {\makecell[l]{\quad \# of Epochs for Training}} &  \makecell[c]{$100$}
        \\
        \midrule
        {\makecell[l]{\quad Batch size}} &  \makecell[c]{$64$}
        \\
        \midrule
        {\makecell[l]{\quad Best-of-N}} &  \makecell[c]{$[4,6,8]$}
        \\
        \midrule
        {\makecell[l]{\quad TopK}} &  \makecell[c]{$[10,15,20]$}
        \\
        \midrule
        {\makecell[l]{\quad TopP}} &  \makecell[c]{$[0.85,0.9,0.95]$}
        \\
        \midrule
        {\makecell[l]{APO Objective Weight}}
        \\
        \midrule
        {\makecell[l]{\quad Docking Score}} &  \makecell[c]{$0.2$}
        \\
        \midrule
        {\makecell[l]{\quad Druglikeliness}} &  \makecell[c]{$0.2$}
        \\
        \midrule
        {\makecell[l]{\quad Synthesizability}} &  \makecell[c]{$0.2$}
        \\
        \midrule
        {\makecell[l]{\quad Solubility}} &  \makecell[c]{$0.2$}
        \\
        \midrule
        {\makecell[l]{\quad Tamimoto Similarity}} &  \makecell[c]{$0.2$}
        \\
        \midrule
        {\makecell[l]{APO Other}}
        \\
        \midrule
        {\makecell[l]{\quad Fingerprint Size}} &  \makecell[c]{$1024$}
        \\
        \midrule
        {\makecell[l]{\quad Normalize Min/Max}} &  \makecell[c]{$[-10, 10]$}
        \\
        \midrule
        {\makecell[l]{Advantage preference with \\ invalid generated SMILES}}
        \\
        \midrule
        {\makecell[l]{\quad 3CLPro}} &  \makecell[c]{$[0,-R_c(X)]$}
        \\
        \midrule
        {\makecell[l]{\quad RTCB}} &  \makecell[c]{$[0,-R_c(X)]$}
        \\

        \bottomrule
    \end{tabular}}}
        \caption{{{Hyperparameters for APO}}. }
        \label{app:tab:hyperparams}
        }
\end{table*}

\end{document}